\documentclass[prd,floats,floatfix,showpacs,nofootinbib]{revtex4}
\usepackage{graphicx}
\usepackage{graphics}
\usepackage{amssymb}
\usepackage{amsmath}
\usepackage{url}
\usepackage{color}
\newcommand{\fracc}[2]{\frac{\textstyle{#1}}{\textstyle{#2}}}

\newcommand{\p}{\partial}

\begin{document}

\title{Dynamical Wormhole Definitions Confronted}

\author{E. Bittencourt$^1$}\email{bittencourt@unifei.edu.br}
\author{R. Klippert$^1$}\email{klippert@unifei.edu.br}
\author{G. B. Santos$^2$}\email{grasiele.santos@unifal-mg.edu.br}

\affiliation{$^1$Universidade Federal de Itajub\'a, Av. BPS 1303, Itajub\'a-MG, 37500-903, Brazil}
\affiliation{$^2$Universidade Federal de Alfenas, Campus Po\c{c}os de Caldas, Rodovia Jos\'e Aur\'elio Vilela 11999, Po\c{c}os de Caldas - MG, 37715-400, Brazil}

\pacs{}
\date{\today}

\begin{abstract}
Crude comparison between four alternative proposals for the very definition of a wormhole is provided, all of which were intended to apply to the dynamical cases. An interesting dynamical solution, based upon large scale magnetic fields, is used for the comparisons. Such solution goes beyond the perfect fluid approximation due to an anisotropic pressure component, bringing to the fore some unsuspected features of those definitions. Certain notions as reversible traversability are claimed as a way to select among those definitions the best suited one to represent our intuition of what a wormhole solution is expected to be.
\end{abstract}

\maketitle

\section{Introduction}

Wormhole solutions have been widely studied in a variety of contexts mainly since the important work of Morris and Thorne \cite{mor88,mor88b} for the static spherically symmetric case, although the first solutions date back to fifteen years earlier \cite{ellis73,bro73}. Loosely speaking, a wormhole is a structure that works as a short cut between two different regions in spacetime or possibly two different universes (for a review, see Ref.\ \cite{lobo}). In particular, it was established that violation of the null energy condition (NEC) would always take place, both in the static and dynamic cases. In the static case, the NEC violation was established using global arguments \cite{visser96} while in the dynamical case quasi-local \(2+2\) definitions of wormholes were put forward by Hochberg and Visser \cite{hoch98a,hoch98b} and by Hayward \cite{hay94,hay99}, with the help of non-operational ingredients as causality in the latter. Within this dynamical context, the wormhole was defined in terms of closed two-dimensional spatial hypersurfaces such that one of the two future-directed null geodesic congruences orthogonal to it is just beginning to diverge (the so called flare-out condition).

In particular, the conformally expanding Morris-Thorne geometry as a natural dynamical extension of the static Morris-Thorne wormhole was introduced in \cite{kar94} (see also \cite{kim96}) with the use of a Friedmann-like spacetime. However, from those works, it was not clear which were the regions connected by the wormhole or how to treat the traversability issue\footnote{Also, as noted in \cite{hoch98a}, the attempt of using embedding techniques usually applied in the static case for the dynamical one, as done in \cite{kar94,kim96}, can be misleading when checking the flare-out condition.}. In \cite{hoch98b} it was explicitly stated that two coordinate patches would be necessary in this case, and that would correspond to two different universes being connected by the wormhole.

It was later claimed \cite{mae09} that the above formal definitions could not include ``cosmological wormholes'' due to the presence of the initial singularity and thus the absence of a past null infinity in these geometries. A new definition was established in order to construct such solutions where complicated matching procedures had to be performed. At the same time, a more intuitive and manageable definition of a wormhole in the spherical symmetric case was introduced by Hayward \cite{hay09}. Few years later, an hybrid definition for a cosmological wormhole was proposed \cite{tomi15} in order to avoid the strong dependence on the choice of the surface foliation present in \cite{mae09}.

More recently, it has been proposed from a purely cosmological point of view \cite{bitt14} a spacetime with constant spatial curvature and a non-null Weyl tensor for which the line element takes the form of a conformally expanding Morris-Thorne one. In this case, the Weyl tensor is related to an anisotropic pressure component in the energy-momentum tensor sourced by a primordial stochastic magnetic field satisfying Maxwell's theory. In the present paper, the initial singularity of that spacetime is removed due to the inclusion of nonlinear contributions from the magnetic field \cite{deLorenci02}, and the resulting solution is interpreted as a cosmological wormhole that connects the two asymptotic Friedmann-like regions. In Sec.~\ref{solution} we present the solution proposed in \cite{bitt14} with this nonlinear matter content in the context of a wormhole geometry and we show explicitly how null geodesics are able to cross from one asymptotic region to another. The quasi-local definitions of a wormhole are briefly reviewed in Sec.~\ref{definitions}, all of which are applied to the above mentioned solution. After a short discussion of the energy conditions and the issues regarding their possible violation in Sec.~\ref{energy}, the four definitions are finally contrasted in Sec.~\ref{two-way}, thus allowing us to suggest which among them would be the most suitable definition of a wormhole.

\section{The nonsingular wormhole solution}\label{solution}

Due to our limitations as local observers of the Universe, the conclusions one can obtain from the current observational data ought to be carried carefully. For instance, the interpretation of the data according with the cosmological principle (homogeneity and isotropy) provides good agreement between the FLRW models and the observations only if dark energy and dark matter are taken into account. The assumptions we shall keep here are the isotropy at large scales and the constancy of the spatial curvature, giving up homogeneity. In this way, let us consider the line element of a time-dependent isotropic space-time with constant spatial curvature
\begin{equation}
\label{fried}
ds^2=dt^2-a^2(t)[d\chi^2+\sigma^2(\chi)d\Omega^2],
\end{equation}
where $t$ represents the cosmic time, $a(t)$ is the scale factor, $d\Omega^2=d\vartheta^2+\sin^2\vartheta d\phi^2$ is the square of the infinitesimal element of solid angle and $\sigma(\chi)$ is the radial function which satisfies
\begin{equation}
\label{cons_R}
2\sigma\,\sigma_{\chi\chi}+\sigma_{\chi}^2+3\epsilon\,\sigma^2=1,
\end{equation}
where $\sigma_\chi= d\sigma/d\chi$, $\sigma_{\chi\chi}= d^{\,2}\sigma/d\chi^2$ and $\epsilon=0,+1,-1$ indicates the sign of the spatial curvature \cite{bitt14}.

We assume as source the nonlinear Lagrangian as an extension of Maxwell's theory of electromagnetism given by
\begin{equation}
\label{non_lin_elec}
L=-\fracc{1}{4}\,F+\alpha\, F^2
\end{equation}
where $F=\,F^{\mu\nu} F_{\mu\nu} = 2 (B^2-E^2)$ and $\alpha$ is an arbitrary parameter that should be positive in order to obtain a nonsingular solution \cite{deLorenci02}. It is reasonable to believe that this Lagrangian could play an important role in the early universe due to the expected highly nonlinear behaviour of the primordial plasma. In particular, the nonlinear term added to the Maxwell Lagrangian can be seen as the first order correction due to quantum processes coming from a full theory like Euler-Heisenberg \cite{mg_grasi}.

The energy momentum tensor for this nonlinear Lagrangian is
\begin{equation}
\label{non_lin_elec_tmunu}
T_{\mu\nu}=-4L_F\,F_{\mu}{}^{\lambda}F_{\lambda\nu}-Lg_{\mu\nu},
\end{equation}
which can be decomposed into its irreducible parts with respect to any normalized time-like vector field $V^{\mu}$ as

\begin{eqnarray}
\label{eq_1rho}
\rho&=&-4L_F\,E^2-L,\\[2ex]
\label{eq_2p}
p&=&L+\frac{4}{3}(E^2-2B^2)L_F,\\[2ex]
q^{\alpha}&=&-4L_F\eta^{\alpha}{}_{\beta\mu\nu}V^{\beta}E^{\mu}B^{\nu},\\[2ex]
\label{eq_3pi}
\pi_{\mu\nu}&=&4L_F\left[E_{\mu}E_{\mu} + B_{\mu}B_{\mu} + \frac{1}{3}(E^2+B^2)h_{\mu\nu}\right],
\end{eqnarray}
with $\rho$ as the energy density, $p$ as the isotropic pressure, $q^{\alpha}$ as the heat flow, $\pi_{\mu\nu}$ as the anisotropic pressure and $h_{\mu\nu}=g_{\mu\nu}-V_{\mu}V_{\nu}$ as the projector onto the space orthogonal to $V^{\mu}$. We shall use the vector field $V^{\mu}$ comoving with the cosmic fluid and choose coordinates such that $V^{\mu}=\delta^{\mu}_0$.

In virtue of the special symmetries of the metric\ (\ref{fried}), the electromagnetic field can be considered as source of the gravitational field only if an averaging process is performed (cf.\ Ref.\ \cite{tolman}). The standard procedure for averaging quantities in a 3-spatial hypersurface of constant time marker is to consider a sequence of volumes \(V\) which converge to a limit \(V_o\), with all \(V\) and \(V_o\) belonging to this hypersurface. The limit volume \(V_o\) must be much larger than the typical scale of {\em macroscopic} volumes \(V_m\) over which the random variables largely fluctuate. On the other hand, \(V_o\) should be much smaller than the total volume of the hypersurface (in the eventual cases where this later volume is finite). For general isotropic universes, however, it may prove more appealling to define surface mean values by averaging over the spheres \(S\) which enclose volumes \(V\) with center at the center of spherical symmetry of the universe, with \(S_m\) enclosing \(V_m\) and \(S_o\) enclosing \(V_o\). Thus, the macroscopic cell is a sphere \(S_m\) such that \(\int_{S_m}\vec B\,\sqrt{-\bar g}\,d^2x^\iota=0\) with \(x^\iota=(\vartheta,\,\phi)\) being the angular coordinates and \(\bar g\) the determinant of the induced metric in the sphere \(S_m\). For any quantity \(X\) defined over the hypersurface, its mean value \(\bar X\) is then proposed to be defined as
\begin{equation}
\label{av_def}
\overline{X}\,= \lim_{S \rightarrow S_0} \fracc{1}{S} \int_{S} X\sqrt{-\bar{g}}\,d^2x^\iota.
\end{equation}

Since our aim is the analysis of the primordial plasma, we consider only isotropic magnetic fields and set to zero the electric fields due to the high conductivity of the fluid at the epoch. Consequently, all moments associated to the electric field $E_i$ and all odd moments associated to the magnetic field $B_i$ are zero. The second moment of the magnetic field is given by

\begin{equation}
\label{rel_tol_3}
\overline{B^iB_j}= -\frac{1}{3}B^2h^i{}_j.
\end{equation}
Note that Latin indices $i,j,l,\ldots$ run form $1$ to $3$. Algebraic consistency of this with Eq.~(\ref{eq_3pi}) requires a certain degree of non-Gaussianity at the fourth order moment. We thus set
\begin{equation}
\label{rel_tol_5}
\overline{B^iB_jB^kB_l}=\frac{1}{12}B^4\delta^i_{(j}\delta^k_{l)}-\frac{1}{80\alpha}\pi^{(i}{}_{(j}\delta^{k)}_{l)}.
\end{equation}
where round brackets mean symmetrization, that is, $A_{(\mu\nu)}= A_{\mu\nu}+A_{\nu\mu}$. Other higher moments of even order are also needed, but we shall not bother to state them explicitly since they are not demanded in the following calculations.

The above presented notation is not completely consistent, since the right-hand side of Eqs.~(\ref{non_lin_elec_tmunu})--(\ref{eq_3pi}) should be written under the bar which denotes the mean value discussed above. However, we shall avoid over-notation. The application of the average procedure to the components of $T^{\mu}{}_{\nu}$ leads to the following effective energy momentum (which is the source of the Einstein field equations)
\begin{equation}
\label{Tmunu}
T^\mu{}_\nu=(\rho+p) V^\mu V_\nu-p\,\delta^\mu_\nu+\pi^\mu{}_\nu.
\end{equation}

Under these assumptions, the mean values of Maxwell's equations are identically satisfied, and Einstein equations furnish
\begin{eqnarray}
\dot a&=&\pm\sqrt{\frac{B_0^2}{6a^2}\left(1-\frac{a_b^4}{a^4}\right)-\epsilon},\label{sc_fac}\\[2ex]
\sigma_{\chi}&=&\pm\sqrt{1-\frac{2M}{\sigma}-\epsilon\sigma^2},\label{first_sigma}\\[2ex]
B&=&\frac{B_0}{a^2(t)},\label{B_a}\\[1ex]
\pi^1_1&=&\frac{2M}{a^2\sigma^3},\label{comp_pi}
\end{eqnarray}
with $\pi^2_2=\pi^3_3=-\pi^1_1/2$. The quantities $B_0$ and $M$ are constants, dot means derivative with respect to the time coordinate and $a_b=(8\alpha B_0^2)^{\frac{1}{4}}$ is the value of the scale factor at the bounce. We have also
\begin{equation}
\rho=\frac{B^2}{2}(1-8\alpha B^2),\quad\mbox{and}\quad p=\frac{B^2}{6}(1-40\alpha B^2)
\end{equation}
for the energy density and the isotropic pressure in terms of the magnetic field, respectively.

The line element (\ref{fried}) in spherical coordinates takes the form
\begin{equation}
\label{ds2_r}
ds^2=dt^2-a^2(t)\left(\frac{d\sigma^2}{1-\epsilon \sigma^2-\frac{2M}{\sigma}}+\sigma^2d\Omega^2\right).
\end{equation}
Notice that the spatial sections at constant time are the same as the Schwarzschild-de Sitter ones, with $\sigma$ as the radial marker. In the special case of flat spatial curvature ($\epsilon=0$), Eqs.\ (\ref{sc_fac}) and (\ref{first_sigma}) can be integrated, yielding
\begin{equation}
\label{sol_a_flat}
a(t)=a_b\left[\left(\frac{t}{t_c}\right)^2+1\right]^{\frac{1}{4}},
\end{equation}
where $t_c=\sqrt{12\alpha}$, and
\begin{equation}
\label{sol_sigma}
\chi = \pm\left[ \sqrt{\sigma^2-2M\sigma} + M\ln\left(\frac{\sigma-M+\sqrt{\sigma^2-2M\sigma} }{M}\right)\right].
\end{equation}
The function $\sigma$ is strictly positive for $\chi\in(-\infty,+\infty)$ and it assumes its minimum value when $\chi=0$.

The geodesic motion in the metric (\ref{fried}) can be easily studied  if we rewrite the line element in the conformal time $d\eta=dt/a$:
\begin{equation}
\label{ds2_chi}
ds^2=a^2(\eta)[d\eta^2-d\chi^2-\sigma^2(\chi)d\Omega^2].
\end{equation}

The integration of the geodesic equations in terms of the affine parameter $s$ is straightforward and yields the following vector field tangent to the congruences
\begin{equation}
\label{vec_int_geod}
V^{\mu} = \frac{dx^{\mu}}{ds} = \frac{1}{a^2}\left( \pm\sqrt{E^2+m\, a^2},\, \pm \sqrt{E^2-\frac{\gamma}{\sigma^2}},\, \pm \frac{1}{\sigma^2} \sqrt{\gamma-\frac{L^2}{\sin^2\vartheta}},\, \frac{L}{\sigma^2\sin^2\vartheta}\right),
\end{equation}
with $E$, $L$ and $\gamma$ arbitrary constants. The geodesics are space-like, light-like or time-like according to the value of $m$ ($-1$, $0$ or $+1$, respectively). Each combination of the signs in Eq.\ (\ref{vec_int_geod}) leads to a physically different congruence of curves: past or future directed; ingoing or outgoing; clockwise or anti-clockwise rotation. Whenever $a(t)$, $\sigma(\chi)$ and $\sin(\vartheta)$ are different from zero in the range of their arguments, the components of such vector fields are bounded, indicating that the manifold is geodesically complete under these assumptions.

Thus, let us investigate the causal structure of the manifold endowed with metric (\ref{ds2_chi}) using the Carter-Penrose formalism. We first fix the angular coordinates $(\vartheta,\phi)$ to reduce the metric (\ref{ds2_chi}) to a 2D form:
\begin{equation}
\label{ds2_chi_2d}
ds^2=a^2(\eta)[d\eta^2-d\chi^2].
\end{equation}
Clearly, this line element is conformally equivalent to the 2D Minkowski space-time (in Cartesian coordinates) for $a\neq0$, which is precisely our case of interest, since we are dealing with non-singular cosmological models (see Eq.\ \ref{sc_fac}). This equivalence can also be confirmed by the boundedness of the components of the radial light-like geodesics: $d\eta/ds=\pm E/a^2$ and $d\chi/ds=\pm E/a^2$.

\begin{figure}[!htb]
\centering
\includegraphics[width=6cm,height=6cm]{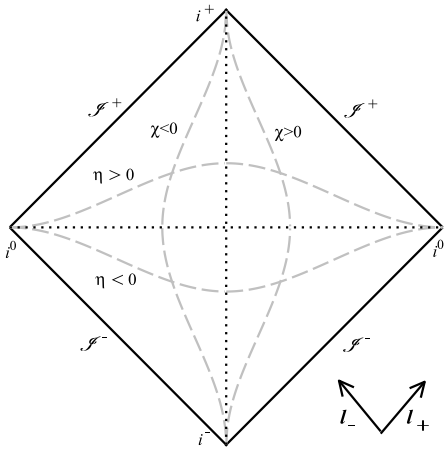}
\caption{Carter-Penrose conformal diagram of the 2-dimensional metric (\ref{ds2_chi_2d}). The symbols $l_+$ and $l_-$ indicate respectively future directed outgoing and ingoing directions on the right-hand side of the diagram, but they have the opposite meaning on the left-hand side.}
\label{conf1}
\end{figure}

Therefore, the same coordinate transformations used to represent the Carter-Penrose conformal diagram of the Minkowski metric can be applied here. That is, we introduce a chart of null coordinates $\sqrt{2}\,u_{\pm}=(\eta\pm\chi)$ and then map the manifold covered by these null coordinates into a finite region of a plane parameterized by $(\psi,\xi)$, for instance, in terms of the coordinate transformation $u_{\pm}=\tan(\psi\pm\xi)$. Fig.~\ref{conf1} depicts such diagram highlighting surfaces of constant $\eta$ and $\chi$. The dotted lines delimit parts of the manifold which are copies of each other: top and bottom, left--hand and right--hand sides.

It should be remarked that the conformal diagram in the big-bang scenario is only the first quadrant of Fig.~\ref{conf1}, where the horizontal dotted line would represent the initial singularity. Furthermore, the conformal diagram of Minkowski space-time in spherical coordinates would be half of Fig.~\ref{conf1}, represented only by the first and fourth quadrants. In virtue of the symmetry of the diagram, the exact solution of Einstein's equations we presented here seems to be a suitable candidate for a wormhole.

\section{Comparison between the dynamical wormhole definitions}\label{definitions}
In this section, we apply the definitions of dynamical wormholes found in the literature and compare their physical interpretation in the light of the non-singular solution presented above. For the sake of simplicity, we shall focus here only on the flat spatial section ($\epsilon=0$) case. The same procedure can be followed for the non-flat spatial section cases ($\epsilon\neq0$) as well as for any non-singular cosmological solution with a spatial metric presenting a non-trivial topology.

\subsection{Hochberg-Visser's definition}\label{defHV}
According to D. Hochberg and M. Visser's (HV) definition \cite{hoch98a,hoch98b}, the wormhole throats $\Sigma_{+}$ and $\Sigma_{-}$ are characterized in terms of sets of null geodesics $l_{+}^{\mu}$ and $l_{-}^{\mu}$, respectively, orthogonal to them. With a $2+2$ decomposition of the manifold, $\Sigma_{\pm}$ (for short) are defined from a variational principle as co-dimension two closed surfaces, the variations of which are performed along arbitrary null directions parameterized by the affine parameters $u_{\pm}$. The conditions for the existence of such minimal area surfaces are equivalent to the vanishing of the expansion coefficients $\Theta_{\pm}= (l_{\pm}^{\mu})_{;\mu}$ of the null congruences under consideration on $\Sigma_{\pm}$ and, besides, the non-negativity of the derivative of $\Theta$ with respect to $u_{\pm}$ (flare-out). Namely, we must have
\begin{eqnarray}
&&\mbox{on}\,\,\Sigma_{+}\,:\quad\Theta_{+}=0 \quad\mbox{and}\quad \frac{d\Theta_{+}}{du_{+}}\geq0,\label{hoch_viss_cond_plus}\\[1ex]
&&\mbox{on}\,\,\Sigma_{-}\,:\quad\Theta_{-}=0 \quad\mbox{and}\quad \frac{d\Theta_{-}}{du_{-}}\geq0.\label{hoch_viss_cond_minus}
\end{eqnarray}
In the case we are dealing with, the radial ($L,\gamma=0$) outgoing and ingoing light-like ($m=0$) vector fields parameterized by the affine parameters in the conformal time satisfying the requirements of the $2+2$ decomposition \cite{carter} are given by
\begin{equation}
\label{vec_int_geod_nul}
l_{\pm}^{\mu}  = \frac{1}{\sqrt{2}\,a^2}\left(1,\,\pm 1,\, 0,\,0\right).
\end{equation}
Thus, the expansion coefficients are
\begin{equation}
\label{theta_viss}
\Theta_{\pm}= (l_{\pm}^{\mu})_{;\mu} = \sqrt{2}\left(\frac{a'}{a}\pm \frac{\sigma_\chi}{\sigma}\right).
\end{equation}
Differentiating these coefficients in terms of the null coordinates $u_{\pm}=(\eta\pm\chi)/\sqrt{2}$, we get
\begin{eqnarray}
&&\frac{d\Theta_{+}}{du_{\pm}} = \frac{a''}{a} - \frac{a'^2}{a^2} \pm \left(\frac{\sigma_{\chi\chi}}{\sigma} - \frac{\sigma_\chi^2}{\sigma^2}\right), \label{diff_theta_viss1}\\[2ex]
&&\frac{d\Theta_{-}}{du_{\pm}} = \frac{a''}{a} - \frac{a'^2}{a^2} \mp \left(\frac{\sigma_{\chi\chi}}{\sigma} - \frac{\sigma_\chi^2}{\sigma^2}\right).
\label{diff_theta_viss2}
\end{eqnarray}
Therefore, all space-time points where the expansion coefficients vanish and the flare-out conditions hold define a wormhole throat, according to the HV definition.

\begin{widetext}
\begin{figure}[!htb]
\centering
\includegraphics[width=5cm,height=5cm]{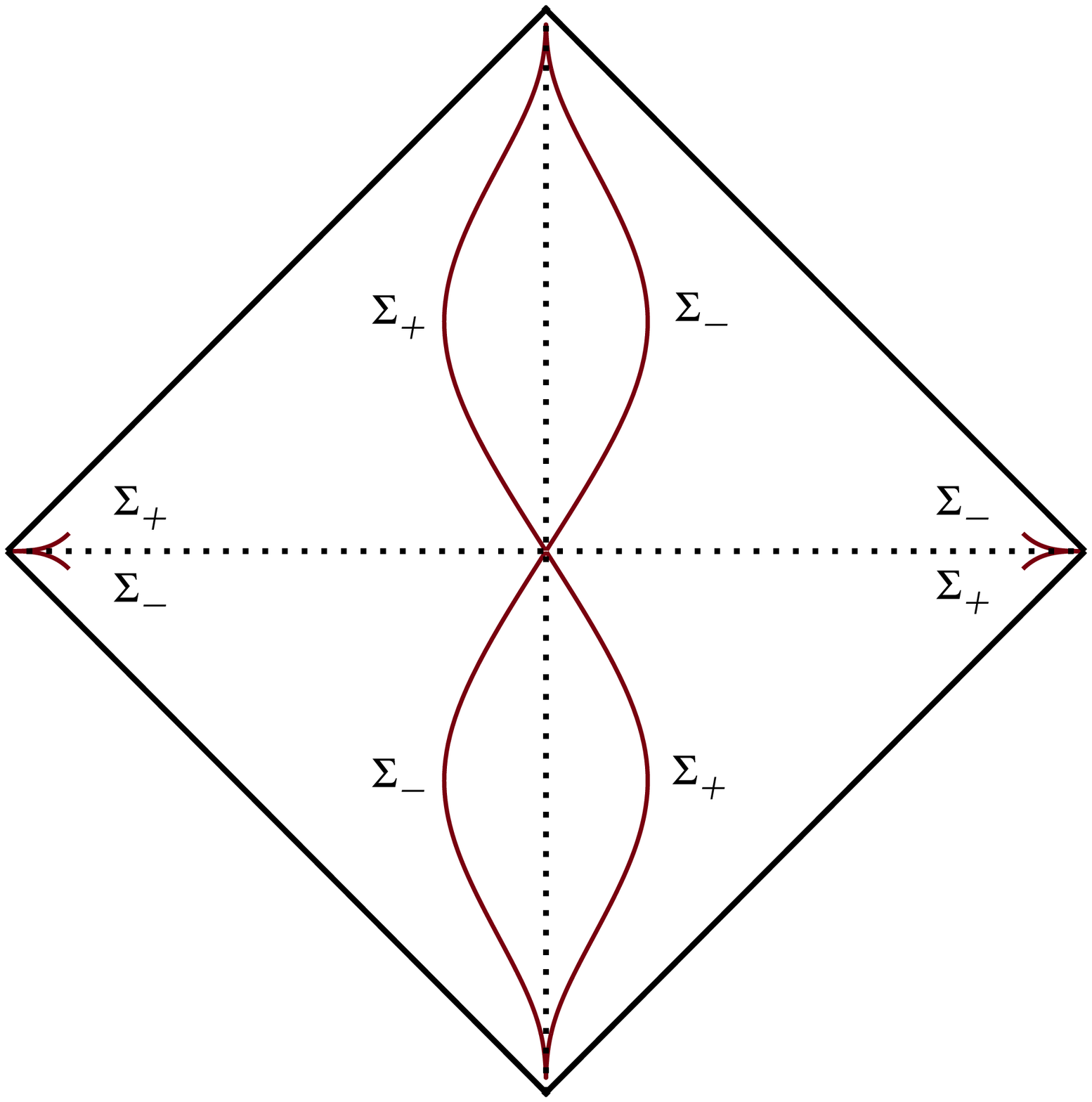}\hspace{.3cm}
\includegraphics[width=5cm,height=5cm]{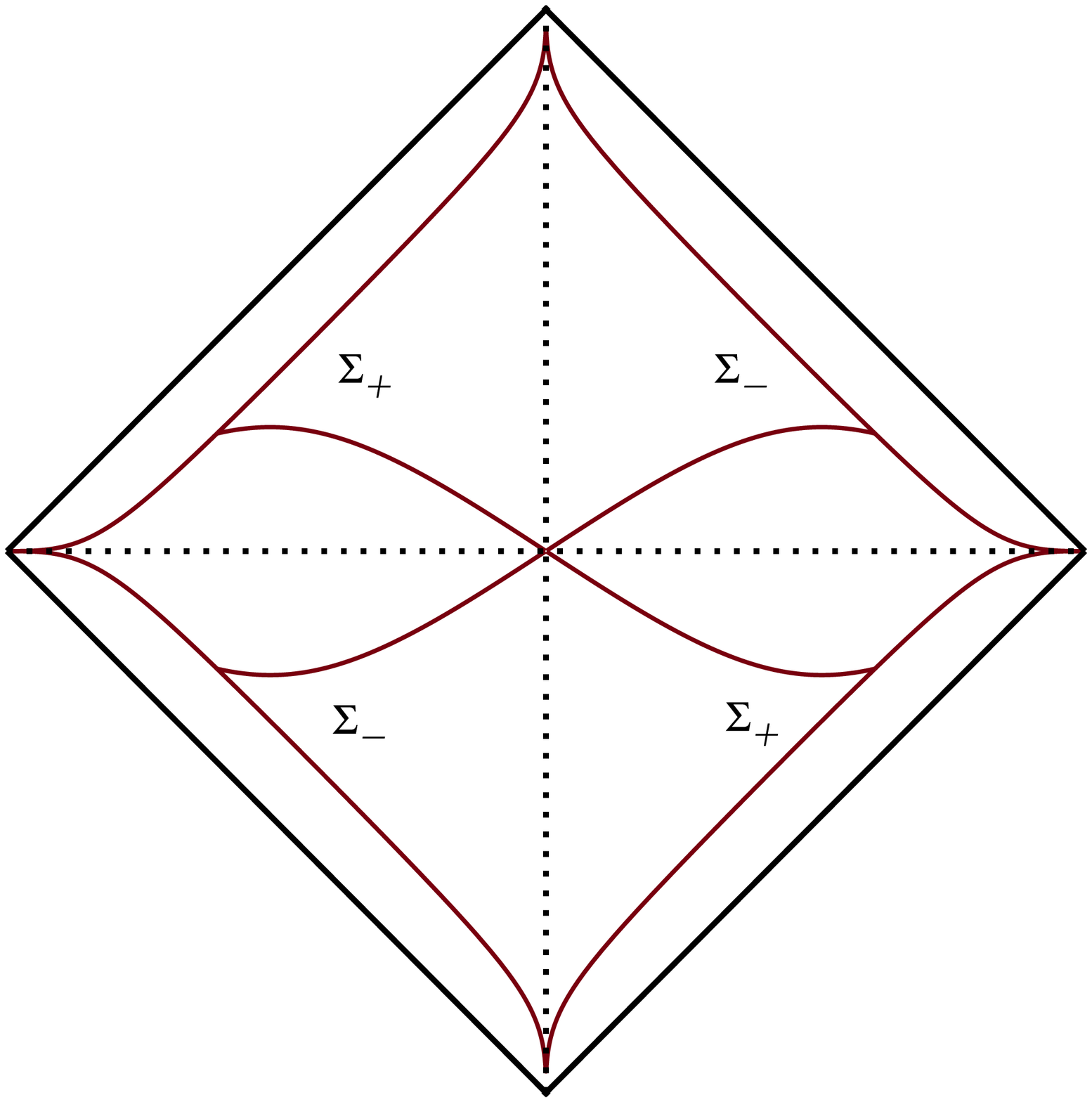}\hspace{.3cm}
\includegraphics[width=5cm,height=5cm]{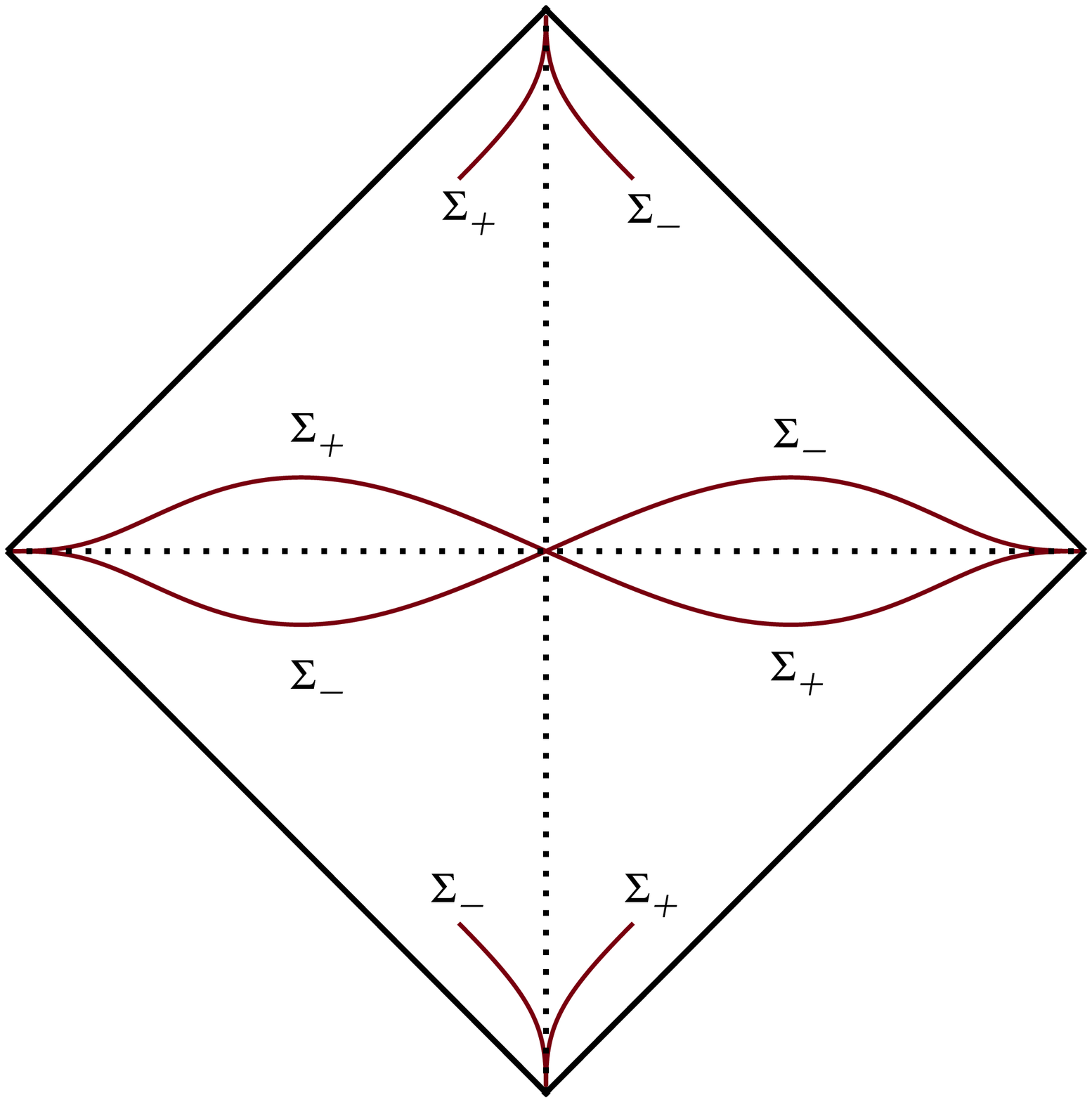}
\includegraphics[width=5cm,height=5cm]{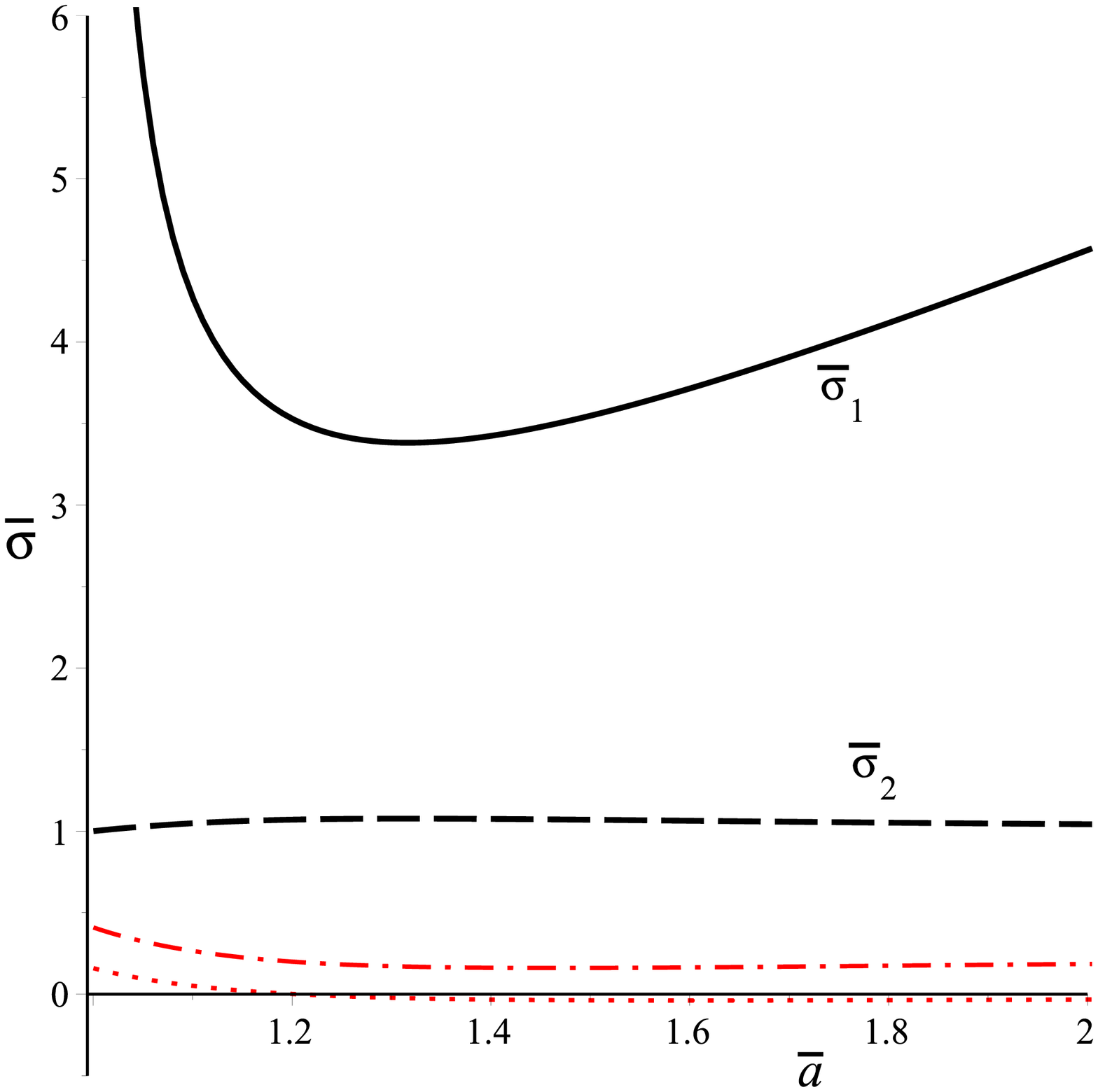}\hspace{.3cm}
\includegraphics[width=5cm,height=5cm]{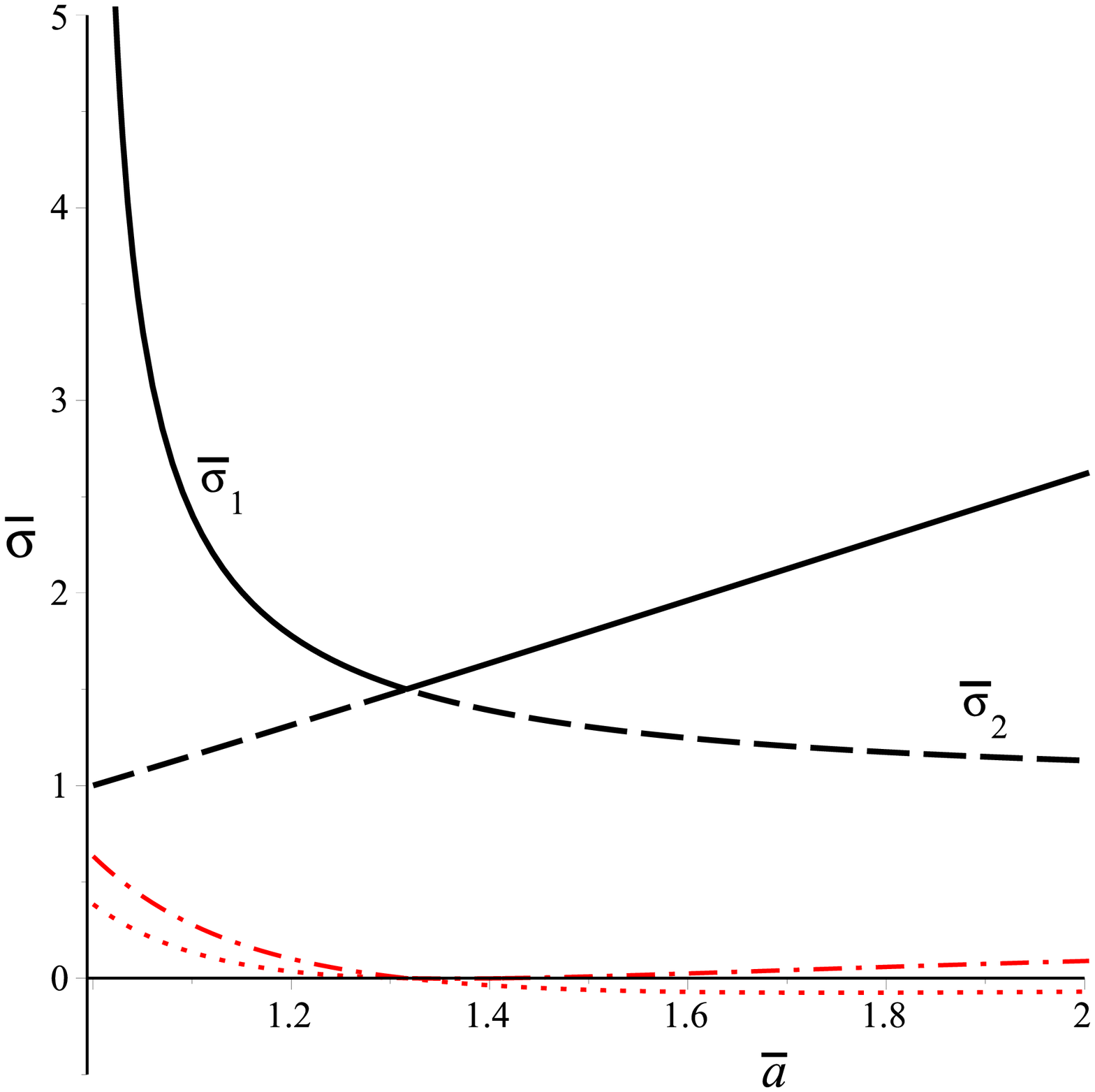}\hspace{.3cm}
\includegraphics[width=5cm,height=5cm]{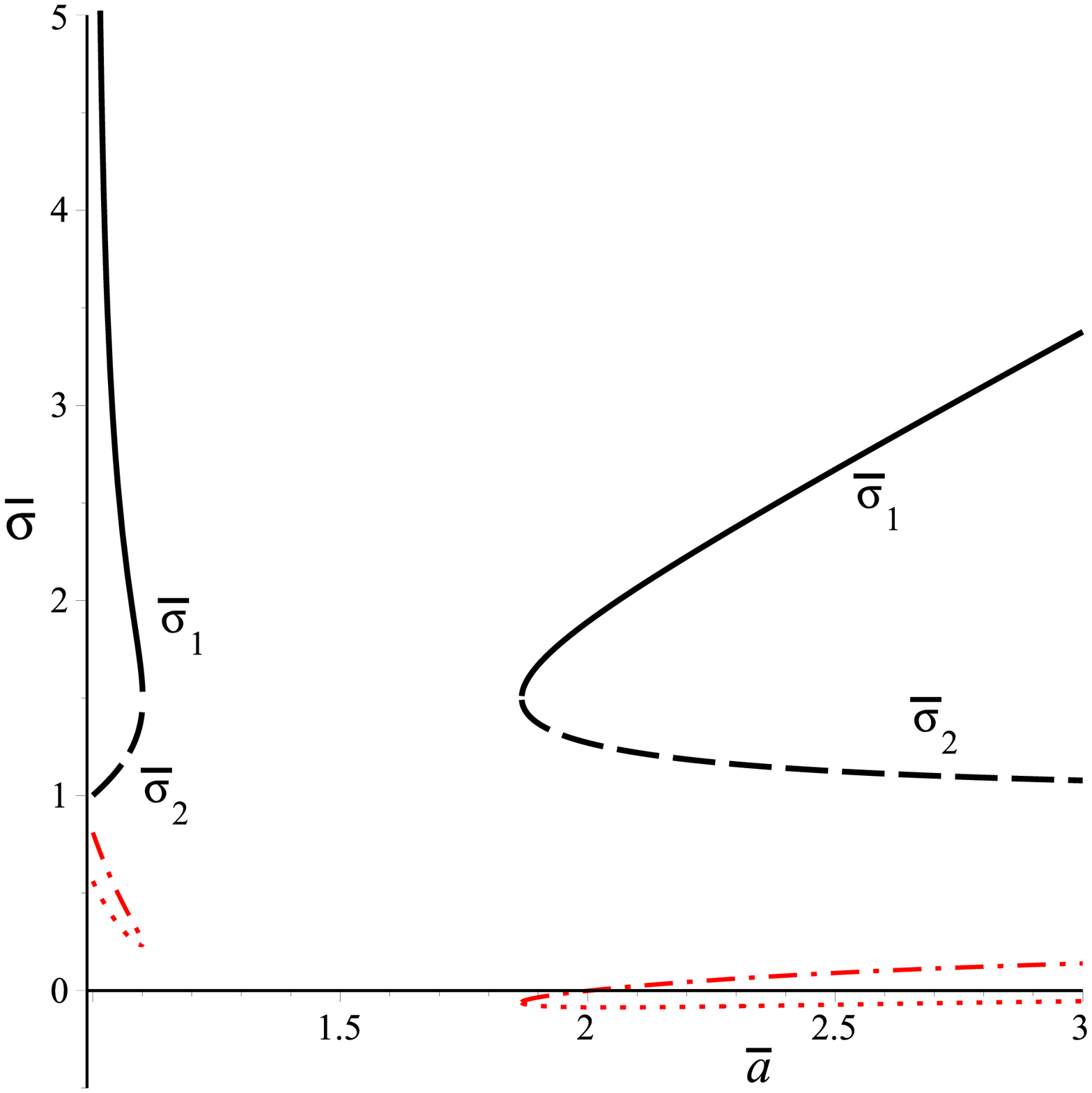}
\caption{Illustrative values of the dimensionless parameter $B_s=0.40$, $B_s=(B_s)_c\approx0.62$ and $B_s=0.75$ are used for the three pairs of panels above in the left, middle, and right, respectively. The panels on the bottom display the solutions of Eq.~(\ref{eq_t_pl2}) identified as \(\bar\sigma{}_1,\,\bar\sigma{}_2\) (the solid and dashed lines, respectively); additionally, the boundary of the flare-out condition [equality at Eq.~(\ref{flare-out})] is also presented as dotted and dash-dotted lines, respectively corresponding to the solid and dashed solutions. The vertical scales do not hold for the flare-outs, since only the algebraic sign of the latter is of any physical significance. The mentioned solutions, with exclusions determined by the corresponding flare-out conditions, are also depicted as the internal lines in the corresponding conformal diagram (the top panels), each point of such internal lines representing a minimal 2-sphere surface at the corresponding time and radius coordinates.}
\label{hv_conf}
\end{figure}
\end{widetext}

Let us first scrutinize the features of $\Sigma_+$. The vanishing of $\Theta_+$ on $\Sigma_+$ implies that
\begin{equation}
\label{eq_t_pl}
\frac{a'}{a}=- \frac{\sigma_\chi}{\sigma}.
\end{equation}
This requires a certain combination of signs between $a'/a$ and $\sigma_\chi/\sigma$, namely, if $a'/a>0$, then $\sigma_\chi/\sigma<0$ and vice-versa. Therefore, this surface should be located at the second and fourth quadrant of the conformal diagram illustrated by Fig.~\ref{conf1}. With the help of Eqs.\ (\ref{sc_fac}) and (\ref{first_sigma}) with $\epsilon=0$ and in the conformal time, we can rewrite Eq.\ (\ref{eq_t_pl}) as
\begin{equation}
\label{eq_t_pl2}
B_s^2(\bar a^4-1)\bar\sigma^3+ (1-\bar\sigma)\bar a^6=0,
\end{equation}
where the barred variables are dimensionless and defined as $\bar\sigma=\sigma/2M$, $\bar a=a/a_b$. We also make use of the dimensionless parameter \(B_s=a_b M/t_c\) as the physical scale of the magnitude of the mean magnetic field. For each value of $\bar a$, the corresponding values of $\bar\sigma=\bar \sigma(\bar a)$ satisfy the cubic equation (\ref{eq_t_pl2}). A simple analysis of this equation shows that it has, at least, one negative real solution for $\bar\sigma$, which consequently must be neglected. The discriminant of (\ref{eq_t_pl2}) is
\begin{equation}
\label{disc_a}
\Delta_{\bar\sigma}=\,B_s^2(\bar a^4-1)\bar a^{12}[4 \bar a^6- 27B_s^2(\bar a^4-1)],
\end{equation}
and its sign determines whether there are other two distinct real roots, one two-degenerate real root, or two complex conjugate roots whenever it is positive, zero, or negative, respectively. The only critical point is located at $\bar \sigma_0=\bar a^3/B_s\sqrt{3(\bar a^4-1)}$ (local minimum). Thus, for each choice of $\bar a$ with $\bar B_s\neq0$ and $\Delta_{\bar\sigma}\geq0$, the other possible values of $\bar \sigma$ are strictly positive. If the discriminant (\ref{disc_a}) vanishes, we get a bi-cubic equation now in terms of $\bar a$:
\begin{equation}
\label{eq_t_pl_sig}
4\bar a^6 - 27B_s^2(\bar a^4-1)=0,
\end{equation}
which, in its turn, has the discriminant given by
\begin{equation}
\label{disc_t_sigma}
\Delta_{\bar a}=4(27)^3B_s^4\left(27B_s^2 - 4\right).
\end{equation}
From this equation, we clearly see that there is a critical value for the scale parameter $(B_s)_c =\sqrt{2/3\sqrt{3}}$ for which the discriminant is zero and the solutions $\bar\sigma (\bar a)$ suffer a bifurcation (see the middle panels in Fig.~\ref{hv_conf}). This cubic equation {in terms of $\bar a^2$} is qualitatively similar to the one for $\bar \sigma$: it has one negative real solution and the value of the critical points imply the positivity of the other two roots whenever $\Delta_{\bar a}\geq0$.

The flare-out condition in this case ($d\Theta_+/du_+\geq0$) requires
\begin{equation}
\label{flare-out}
\bar a^6-4B_s^2(\bar a^4-2)\,\bar\sigma^3\geq0.
\end{equation}
Since the solutions of the cubic equation are known as $\bar\sigma=\bar\sigma(\bar a)$, we then analyze the image of this function, namely, the left-hand side of Eq.\ (\ref{flare-out}), and identify the points in which it is non-negative, for some representative values of $B_s$. The results are presented in Fig.~\ref{hv_conf}. The throats denoted by $\Sigma_-$ can be trivially obtained if one notices that $\Theta_-=0$ and $d\Theta_-/du_-\geq0$ correspond to the same equations as those for $\Sigma_+$ when the first and third quadrants are taken into account. These results are also displayed in Fig.~\ref{hv_conf}.

The conformal diagrams represented in Fig.~\ref{hv_conf} illustrate the position of the wormhole throats in each case, depending on the value of $B_s$. It is worth noticing that, from this definition of wormhole throats, we are able to distinguish which features of the wormhole are emphasized: either its dynamic character (the region near the bounce and far from the center of symmetry), or else its topological character (the regions near the vertical axis of symmetry of the conformal diagram). For instance, for $B_s$ smaller than its critical value, the union of the throats turns out to be defined for all time, forming three connected components, one of which is a borderless time-like line in the conformal diagram. As $B_s$ goes to zero we thus approach the static regime. On the other hand, for $B_s$ greater or equal to its critical value, the throats are inevitable for any observer coming from the past infinity $i^-$ and the middle portion of the throats can be crossed only once due to causality, which has no correspondence to the static case. For the sake of completeness, it should be mentioned that in all space-time points for which the throats are defined the NEC is violated by construction.

\subsection{Hayward's definition}\label{defH}
Contrasting to the previous definition, S. Hayward (SH) elaborates a \(2+2\) definition \cite{hay99,hay09} which is more physically sensible, although not being mathematically well-posed, at a first glance. SH proposed to define wormhole mouths as temporal outer trapping horizons with opposite senses and in mutual causal contact. Particularly, in Ref.\ \cite{hay09} it was explained how to implement mathematically the tools of his definition to the cases of spherically symmetric space-times of the form
\begin{equation}
\label{ds2_hay}
ds^2=2\,e^{2\phi(u_+,u_-)}du_+du_{-} - [r(u_+,u_-)]^2d\Omega^2.
\end{equation}
\begin{widetext}
\begin{figure}[!htb]
\centering
\includegraphics[width=5cm,height=5cm]{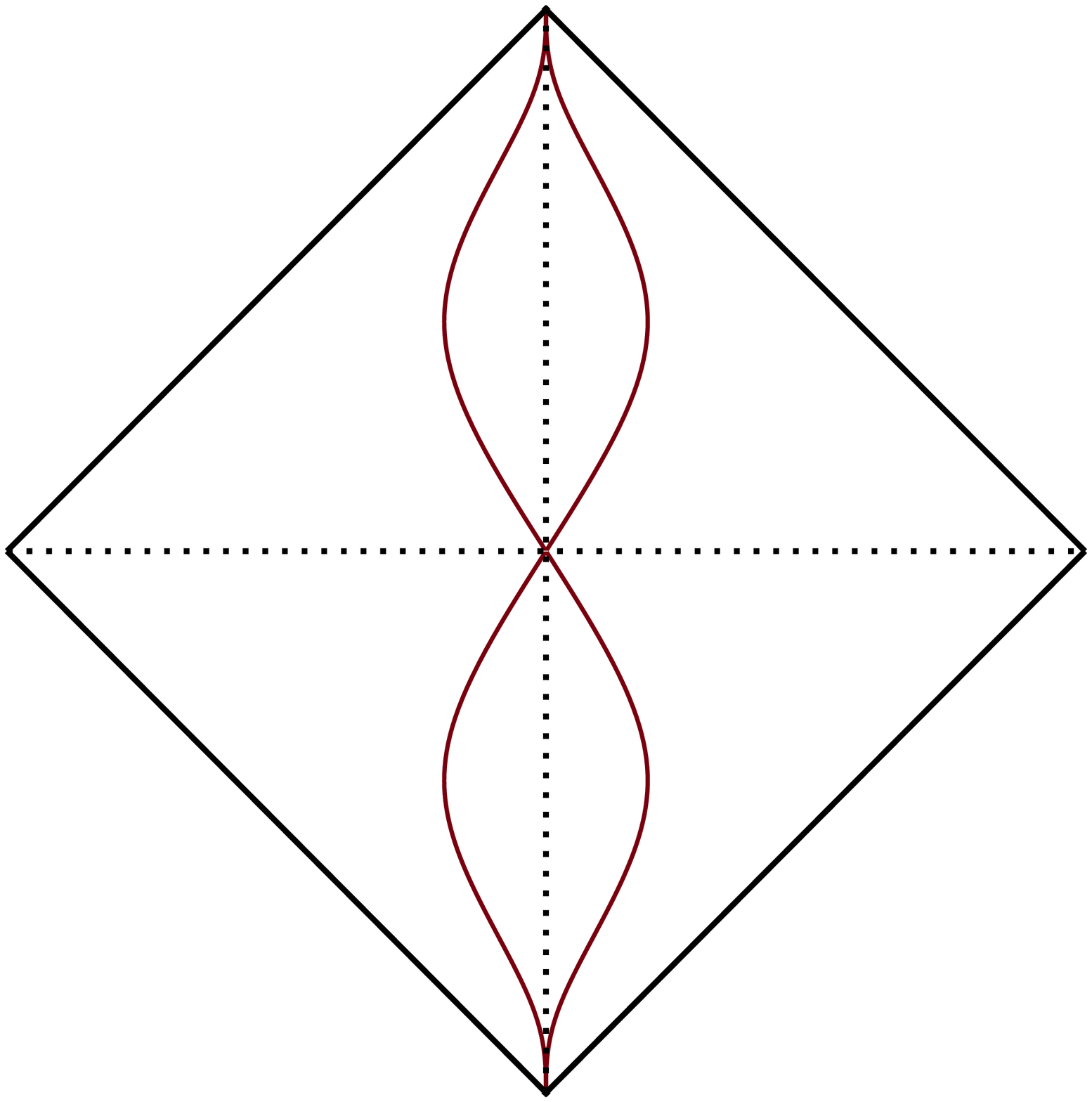}\hspace{.3cm}
\includegraphics[width=5cm,height=5cm]{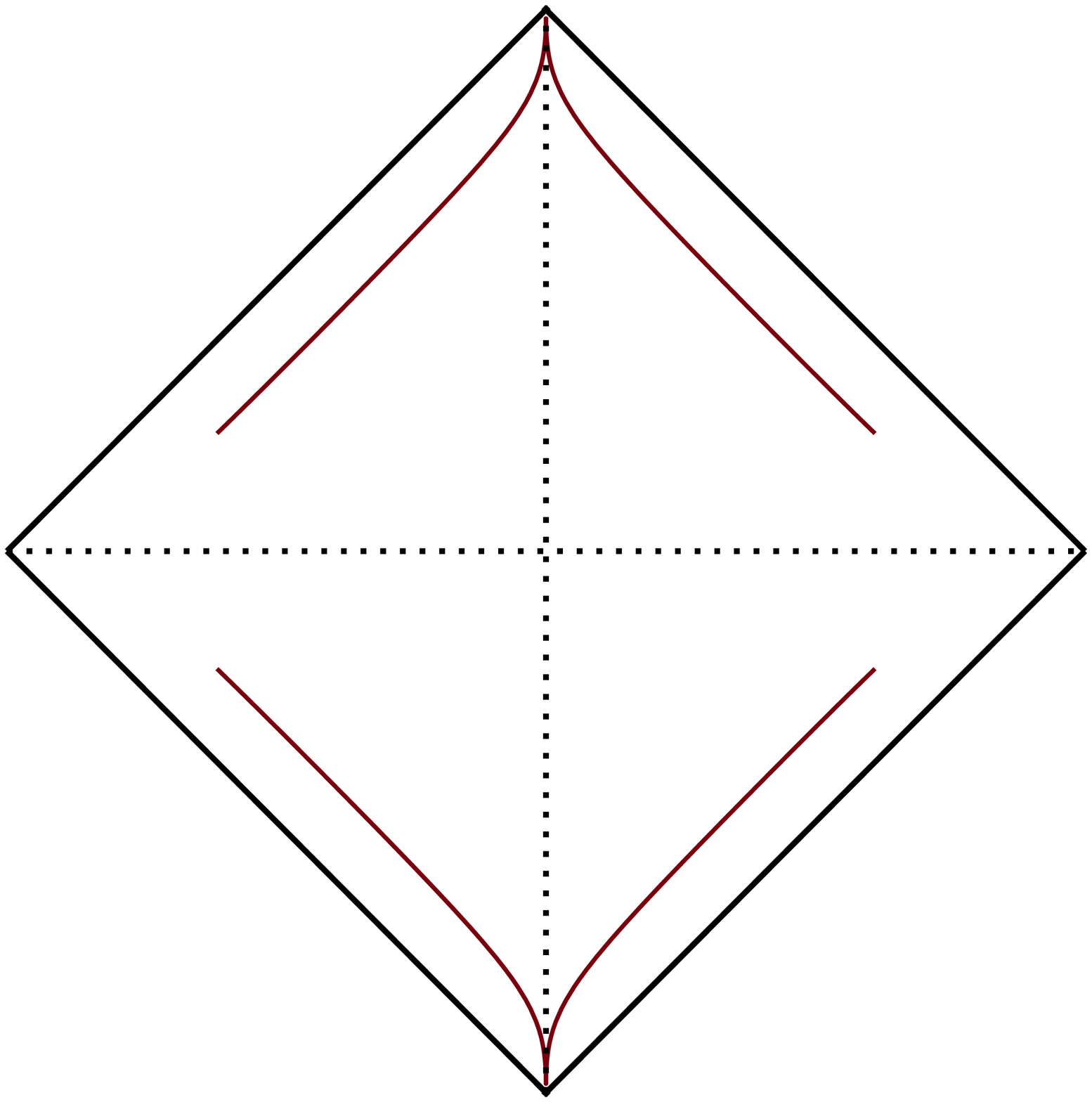}\hspace{.3cm}
\includegraphics[width=5cm,height=5cm]{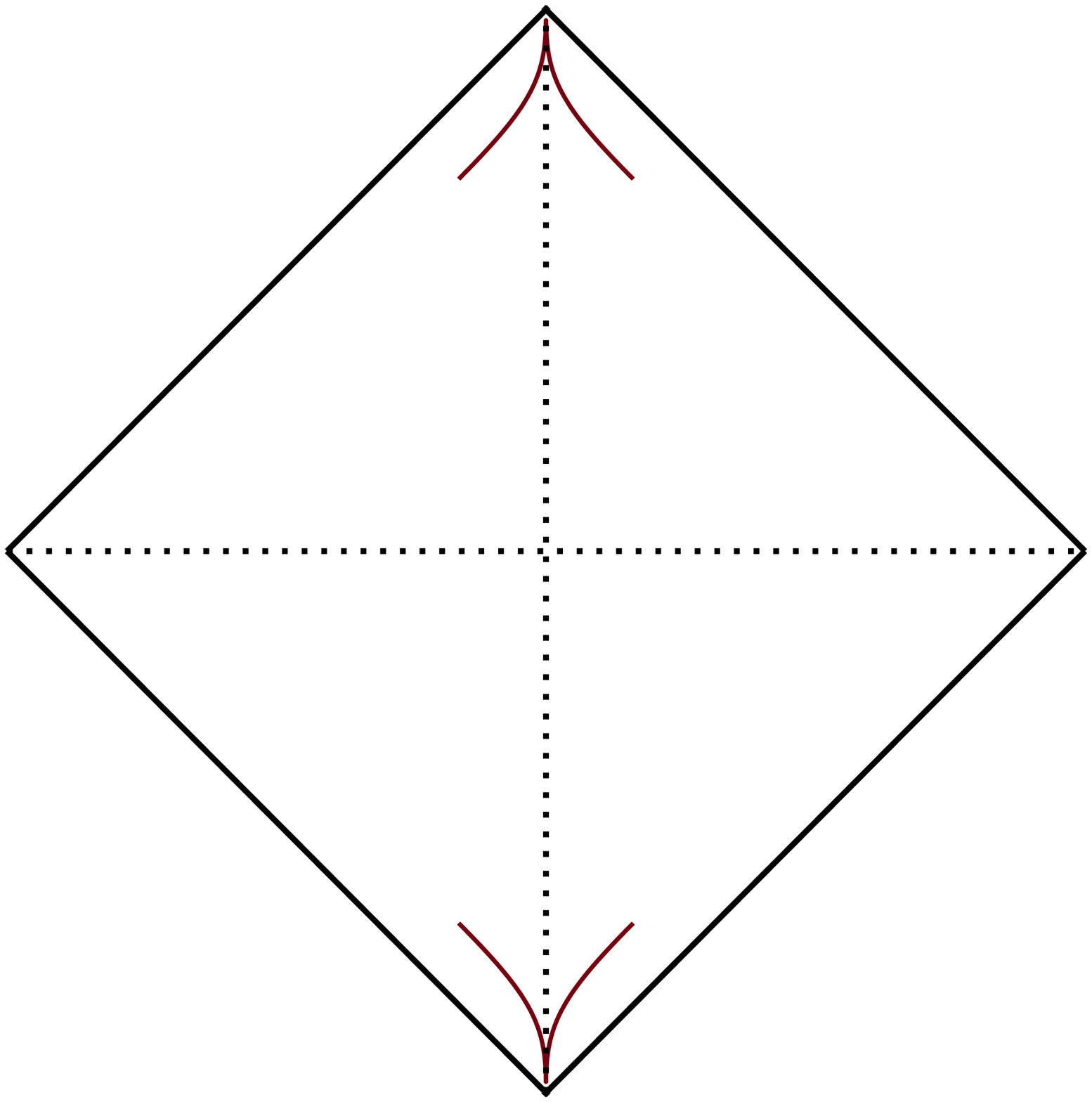}
\includegraphics[width=5cm,height=5cm]{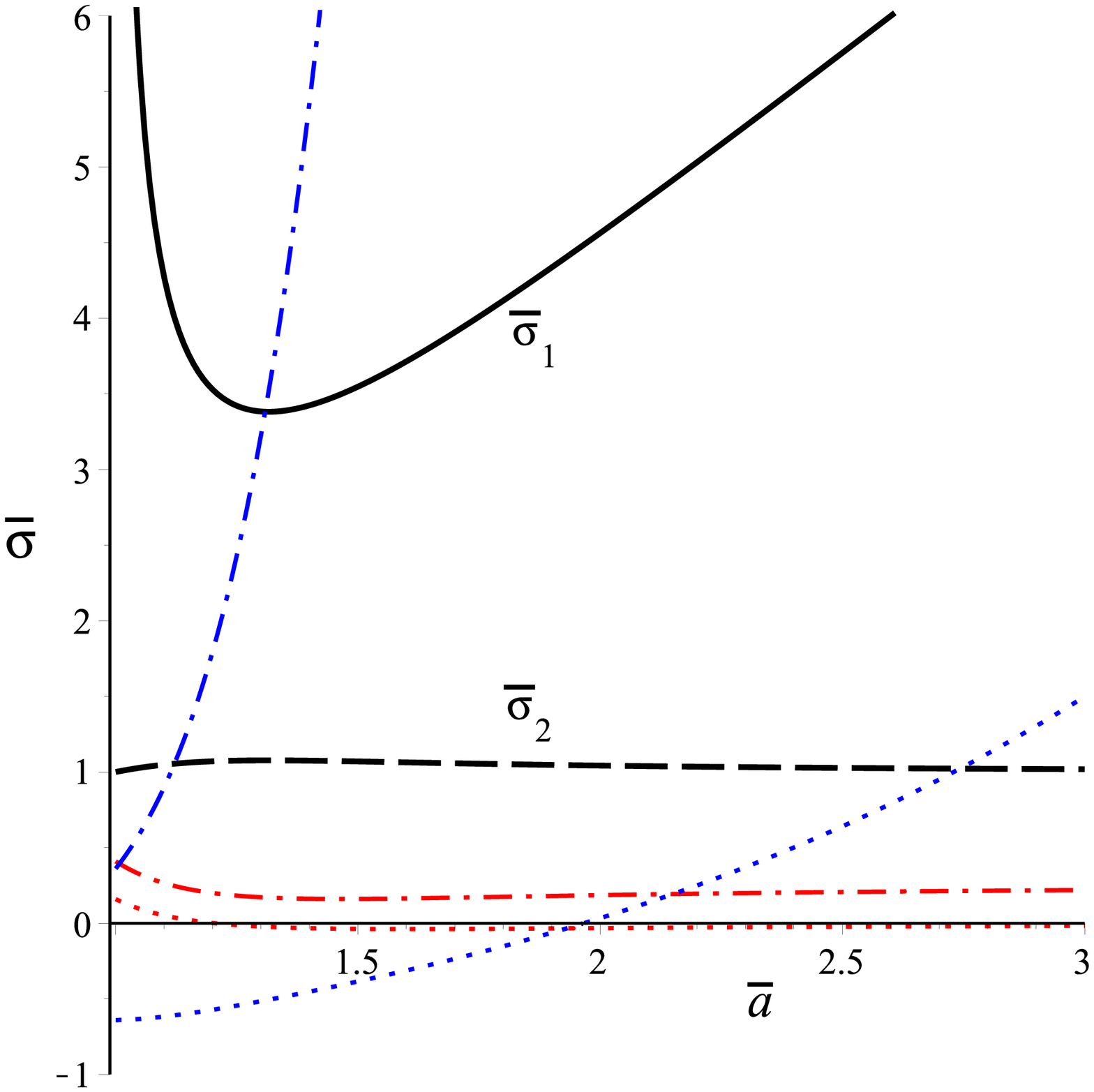}\hspace{.3cm}
\includegraphics[width=5cm,height=5cm]{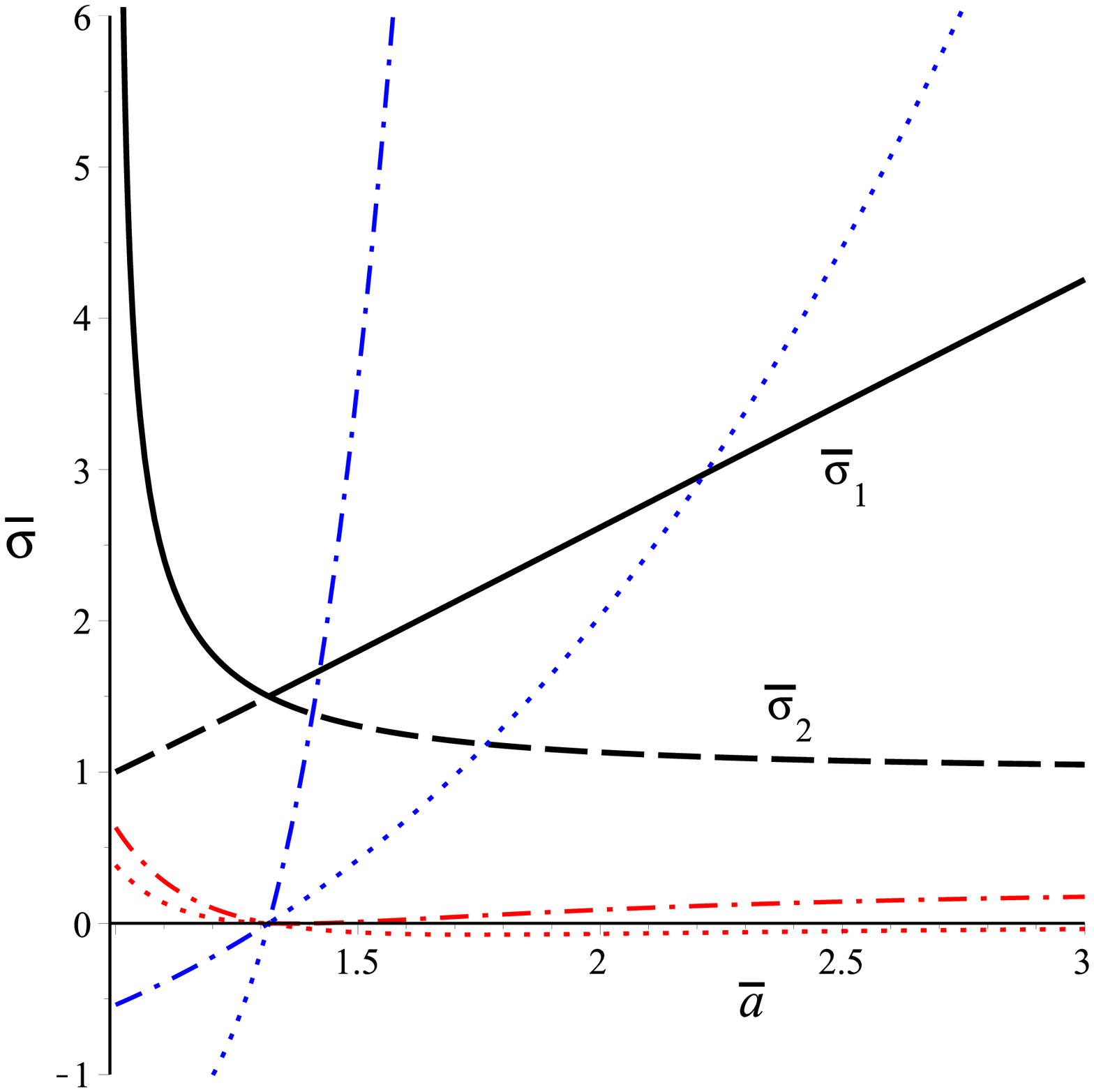}\hspace{.3cm}
\includegraphics[width=5cm,height=5cm]{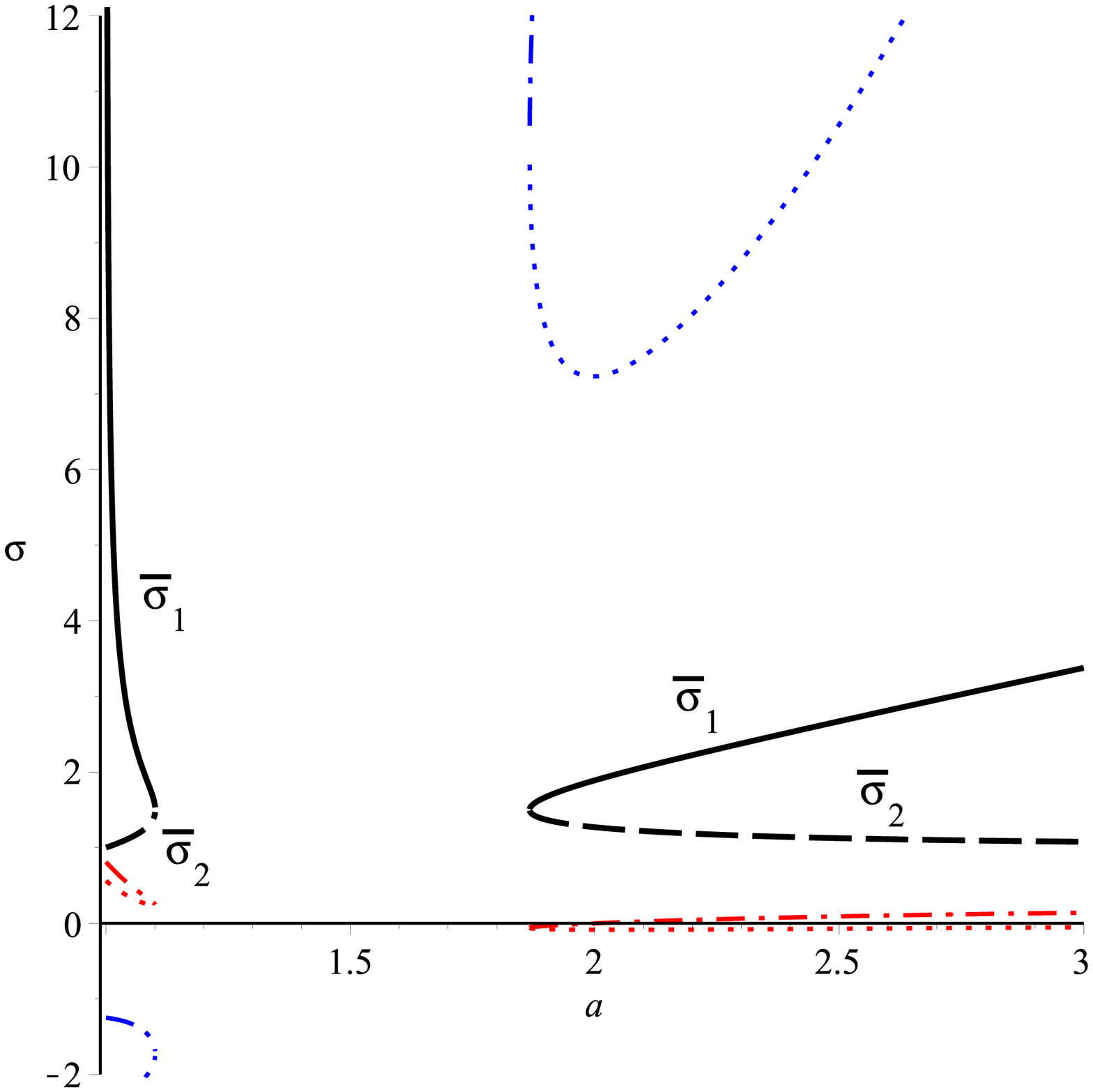}
\caption{Illustrative values of the dimensionless parameter $B_s=0.40$, $B_s=(B_s)_c\approx0.62$ and $B_s=0.75$ are used. On the bottom, the panels display the solutions of Eq.~(\ref{eq_t_pl2}) identified as \(\bar\sigma{}_1,\,\bar\sigma{}_2\) (the solid and dashed lines, respectively); additionally, the boundary of the minimal and outer conditions [equality at Eqs.~(\ref{flare-out}) and (\ref{outer})] are also presented as dotted and dash-dotted lines, respectively corresponding to the solid and dashed solutions. The vertical scales do not hold for the ``flare-outs'', since only their algebraic sign are of any physical significance. The mentioned solutions, with exclusions determined by the corresponding flare-out conditions, are also depicted as the internal lines in the corresponding conformal diagram (the top panels). Each point of such internal lines represents a marginal surface and the line itself is the trapping horizon according to SH's definition.}
\label{hay_conf}
\end{figure}
\end{widetext}

In order for the mouth to be a trapping surface, its radial marker \(r\) should have a minimal character. In terms of the previously considered two classes of null geodesics with parameters \(u_\pm\), extremality condition is expressed either as \(\partial_+r=0\) or \(\partial_-r=0\), while strict minimality condition is expressed respectively as \(\partial_+\partial_+r>0\) or \(\partial_-\partial_-r>0\). Let us assume the \(u_+\) conditions, for definiteness.

The condition for the surface defining the mouth to be of a temporal character can be expressed as \(\xi\cdot d(\partial_+r)=0\), where \(\xi=\xi_+\partial_++\xi_-\partial_-\) is an arbitrary future-directed time-like radial vector, thus implying \(\xi_+>0\) and \(\xi_->0\). The graphical interpretation of this proposal is provided in Fig.~\ref{hay_conf} with similar notation as the previous HV diagrams. However, instead of requiring minimality for the marginal surfaces, SH first demands them to be outer (i.e., $\partial_{\pm}\partial_{\mp}r<0$), and then adding the temporal condition he demonstrates that minimality is a consequence. With this approach, he guarantees the two-way traversibility of the wormhole from the beginning, that is, he assures that it is possible to cross the mouths back and forth. It is straightforward to see that any two among the three hypotheses above (temporal, outer and minimal) are needed; the remaining one turns out to be immediately satisfied. This is helpful for the mathematical implementation of SH's definition.

For the application of this definition in our non-singular solution, the extremality condition is given by Eq.\ (\ref{eq_t_pl2}) and minimality is expressed by (\ref{flare-out}). Therefore, the addition of the outer condition, adapted to our case as
\begin{equation}
\label{outer}
\bar a^6 - 4B_s^2\, \bar\sigma^3\geq0,
\end{equation}
renders SH's definition a particular case of HV's definition. Thus, we construct Fig.~\ref{hay_conf} taking into account the minimal and outer conditions.

\subsection{Maeda, Harada and Carr's definition}\label{defMHC}
In the definition of a cosmological spherically symmetric dynamical wormhole provided by H. Maeda, T. Harada and B. Carr (MHC), the authors also construct the wormhole throats from minimal $2-$spheres \cite{mae09}. In order to do it, they start from the line element Eq.\ (\ref{ds2_hay}) and characterize the wormhole throats by calculating the variation of the volume of the 2-spheres along the null directions tangent to them. That is, the quantities
\begin{equation}
\label{theta_maeda}
\theta_{\pm}= \frac{2}{r}\frac{\p r}{\p u_{\pm}}
\end{equation}
are defined to measure such variation. Note that the null coordinates here are the same as in the previous sections. In Ref.\ \cite{mae09}, the authors claim that a good definition for dynamical wormholes in a cosmological scenario would be taking minimal spheres on space-like hypersurfaces. The conditions for this, on the hypersurface, are
\begin{eqnarray}
&&r_{|A}\zeta^A=0,\label{maeda_cond1}\\[2ex]
&&r_{|AB}\zeta^A\zeta^B>0,\label{maeda_cond2}
\end{eqnarray}
where the indices $A$ and $B$ correspond to the 2D space spanned by $\p_+$ and $\p_-$ and $|$ means covariant derivative on such space. $\zeta^A$ is an arbitrary non-zero radial space-like vector. Eq.\ (\ref{maeda_cond1}) implies that
\begin{equation}
\mbox{either}\quad \theta_+\theta_->0, \qquad \mbox{or} \qquad  \theta_+=\theta_-=0,
\end{equation}
From the line element (\ref{ds2_chi}), the components of the metric can be identified as being $e^{\phi(u_+,u_-)}=a(\eta)$ and $r(u_+,u_-) = a(\eta) \sigma(\chi)$, and we then obtain the following expressions:
\begin{equation}
\label{calc_theta_maeda}
\theta_{\pm}=\sqrt{2}\left(\frac{a'}{a}\pm \frac{\sigma_\chi}{\sigma}\right).
\end{equation}
For any $\zeta^A=\zeta^+\p_{+}+\zeta^-\p_{-}$, with $\zeta^+\zeta^-<0$, the condition (\ref{maeda_cond1}) leads to two distinct cases
\begin{eqnarray}
&&\mbox{(i):}\quad \frac{a'^2}{a^2}>\frac{\sigma_\chi^2}{\sigma^2},\label{maeda_cond1a}\\
&&\mbox{(ii):}\quad \frac{a'}{a}=\frac{\sigma_\chi}{\sigma}=0,\label{maeda_cond1b}
\end{eqnarray}
while the condition (\ref{maeda_cond2}) requires that
\begin{equation}
\label{maeda_cond2c}
\left[\frac{a''}{a}-\left(\frac{a'}{a}\right)^2\right](\zeta^0)^2+\left[\frac{\sigma_{\chi\chi}}{\sigma}-\left(\frac{a'}{a}\right)^2\right](\zeta^1)^2>0,
\end{equation}
where $\zeta^0=(\zeta^{+}+\zeta^{-})/\sqrt{2}$ and $\zeta^1=(\zeta^{+}-\zeta^{-})/\sqrt{2}$ are the radial space-like vector components in the $(\eta,\chi)$ basis, satisfying $(\zeta^0)^2-(\zeta^1)^2<0$. Note that the position of the throats do not coincide with the ones in either HV's or SH's definitions in general. The only case where they coincide is the case (ii) which occurs at the minimum radius at the bounce. The contour of the region delimited by case (i) can be obtained assuming the equality between right-hand and left-hand sides of Eq.\ (\ref{maeda_cond1a}). This leads precisely to the same expression for the extremality conditions in HV's and SH's definitions, which is provided by Eq.\ (\ref{eq_t_pl2}) with bifurcation at the free parameter $B_s$.

\begin{widetext}
\begin{figure}[!htb]
\centering
\includegraphics[width=5cm,height=5cm]{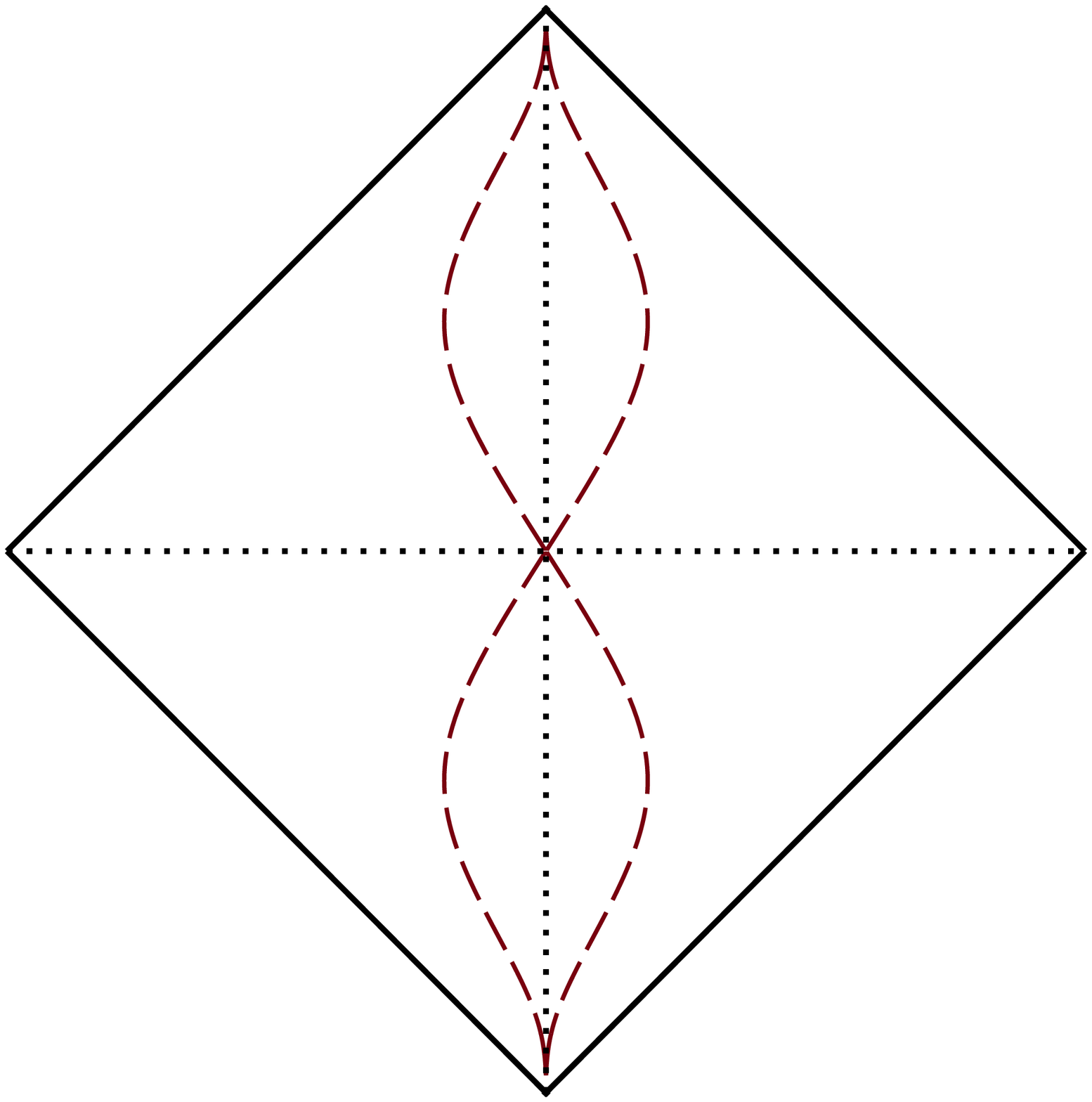}\hspace{.3cm}
\includegraphics[width=5cm,height=5cm]{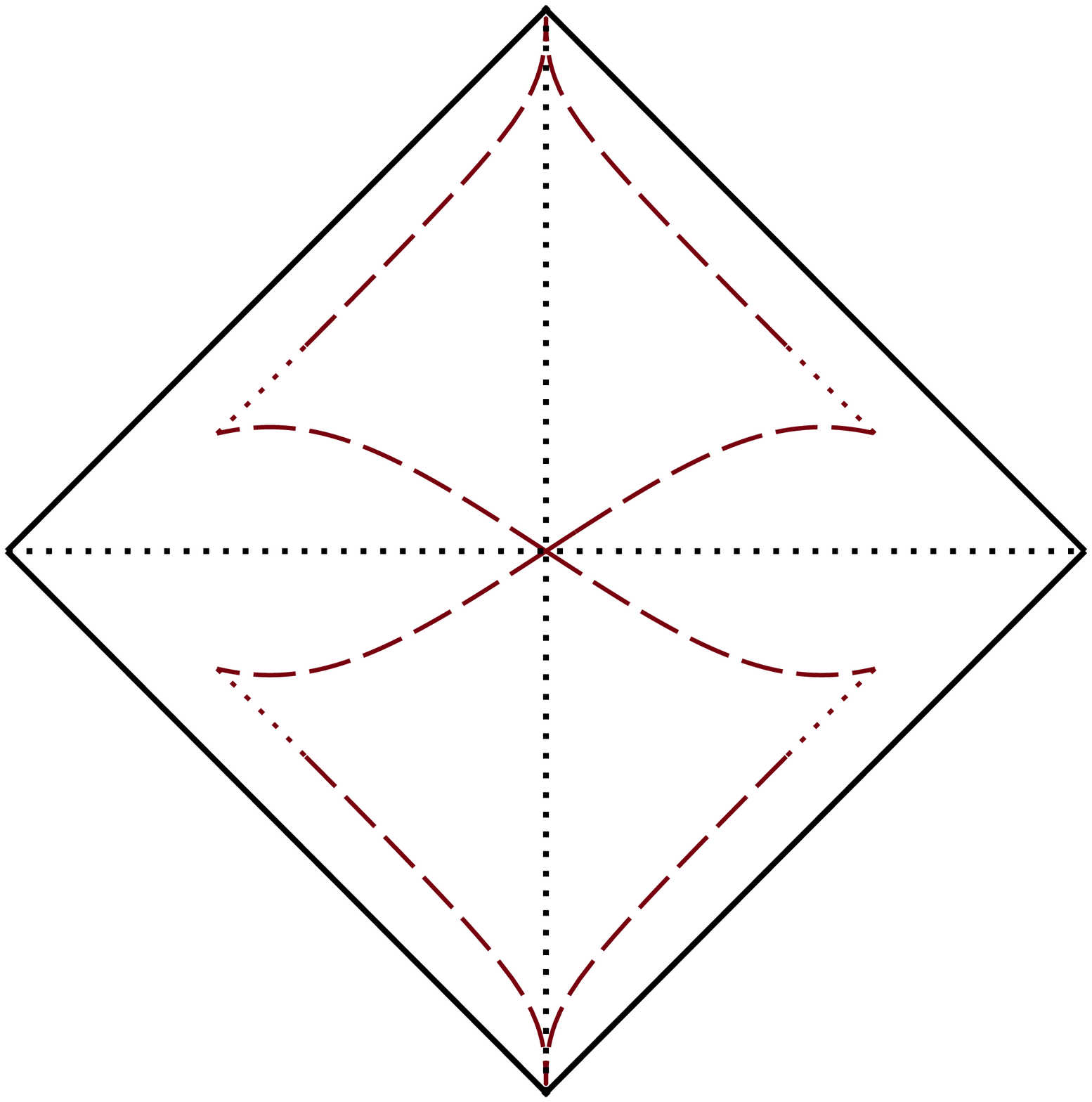}\hspace{.3cm}
\includegraphics[width=5cm,height=5cm]{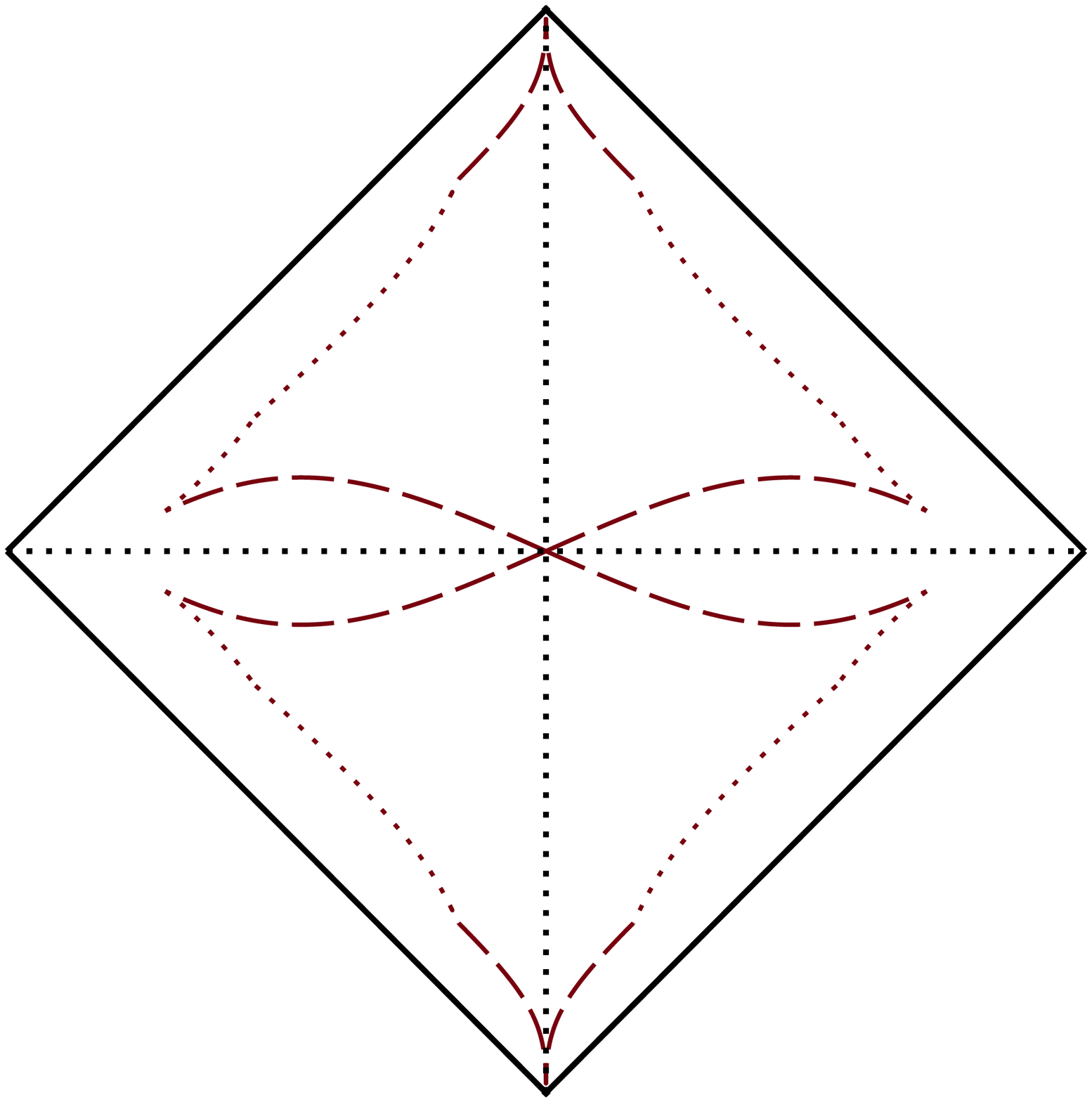}
\includegraphics[width=5cm,height=5cm]{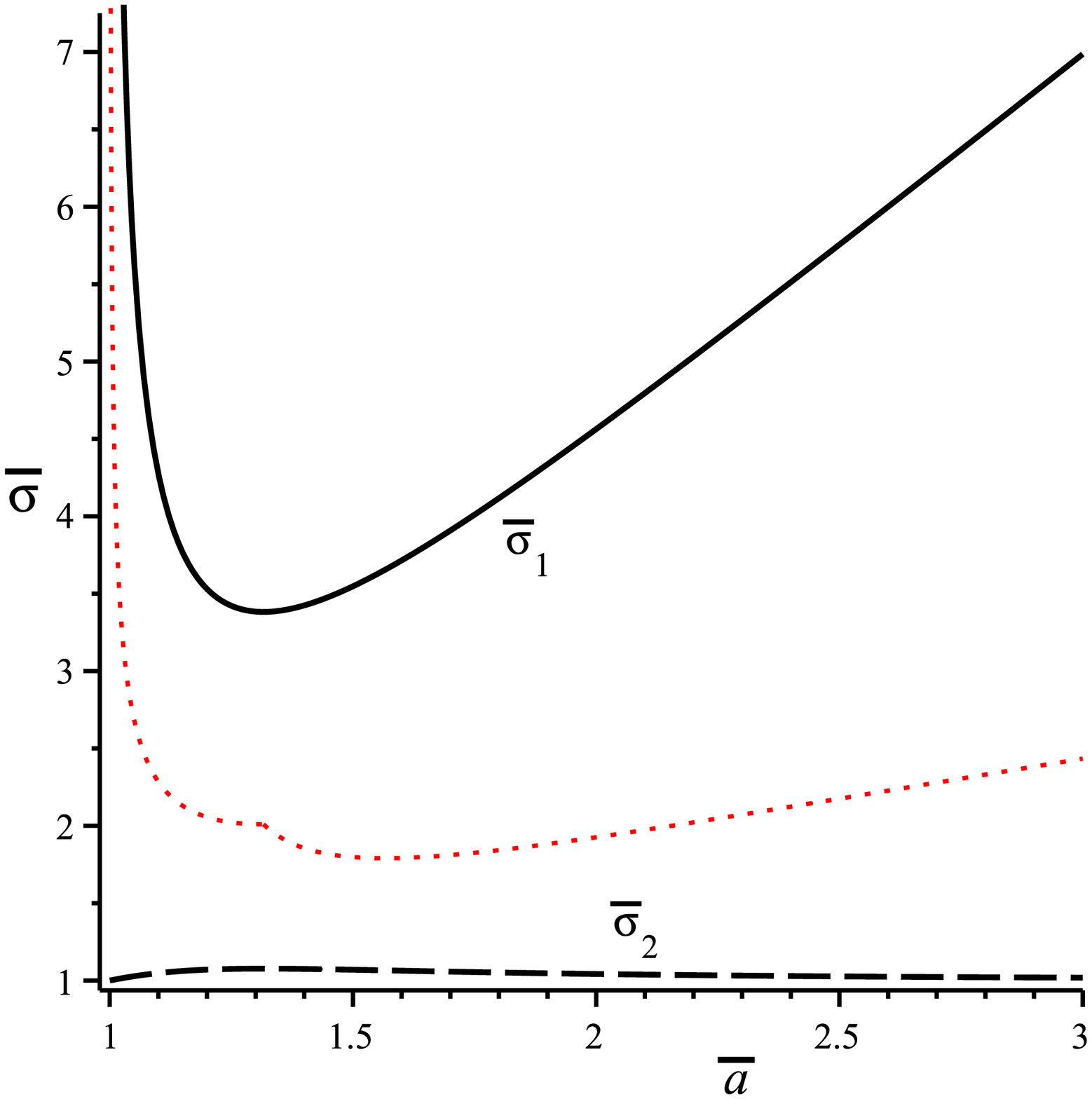}\hspace{.3cm}
\includegraphics[width=5cm,height=5cm]{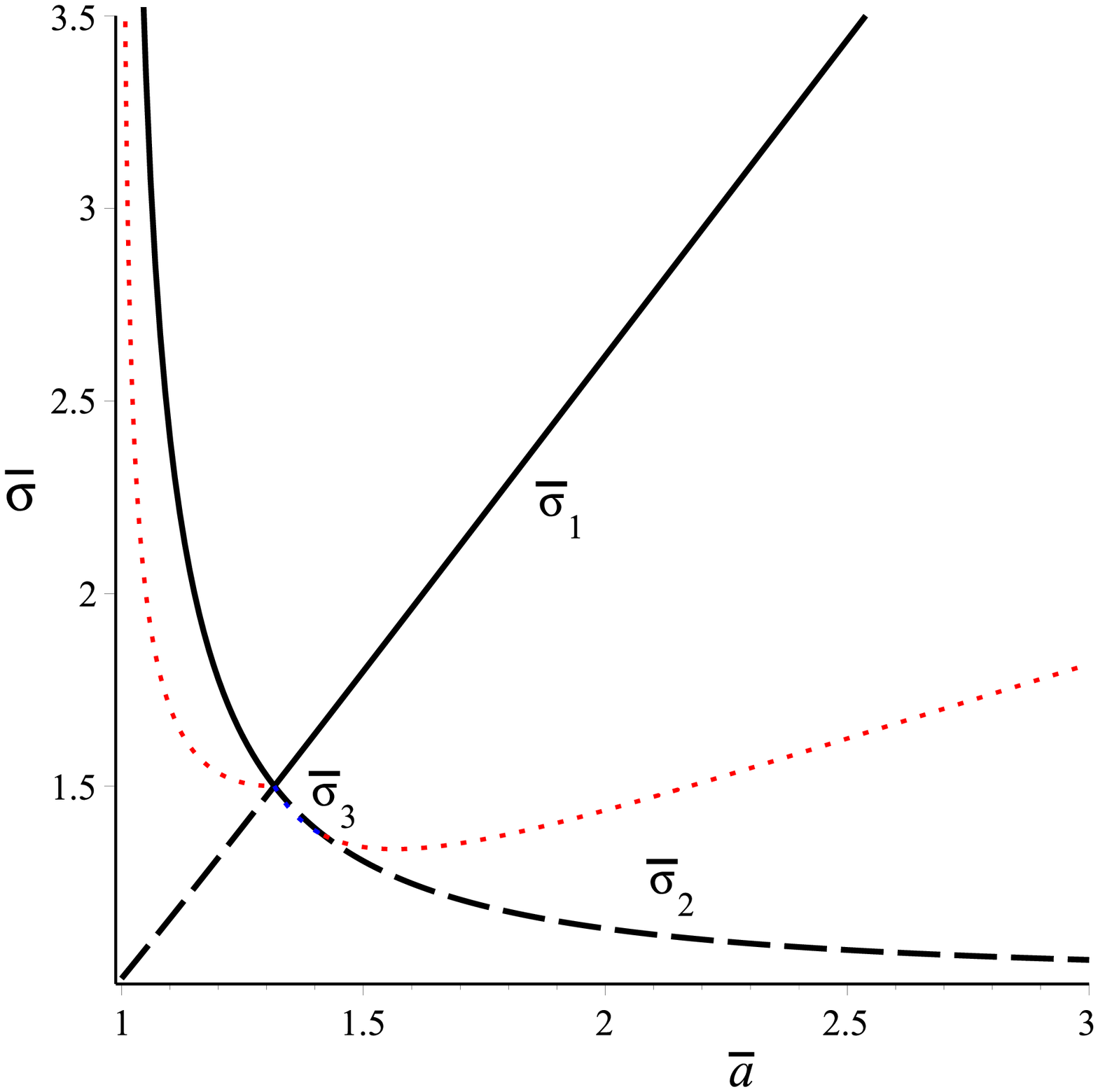}\hspace{.3cm}
\includegraphics[width=5cm,height=5cm]{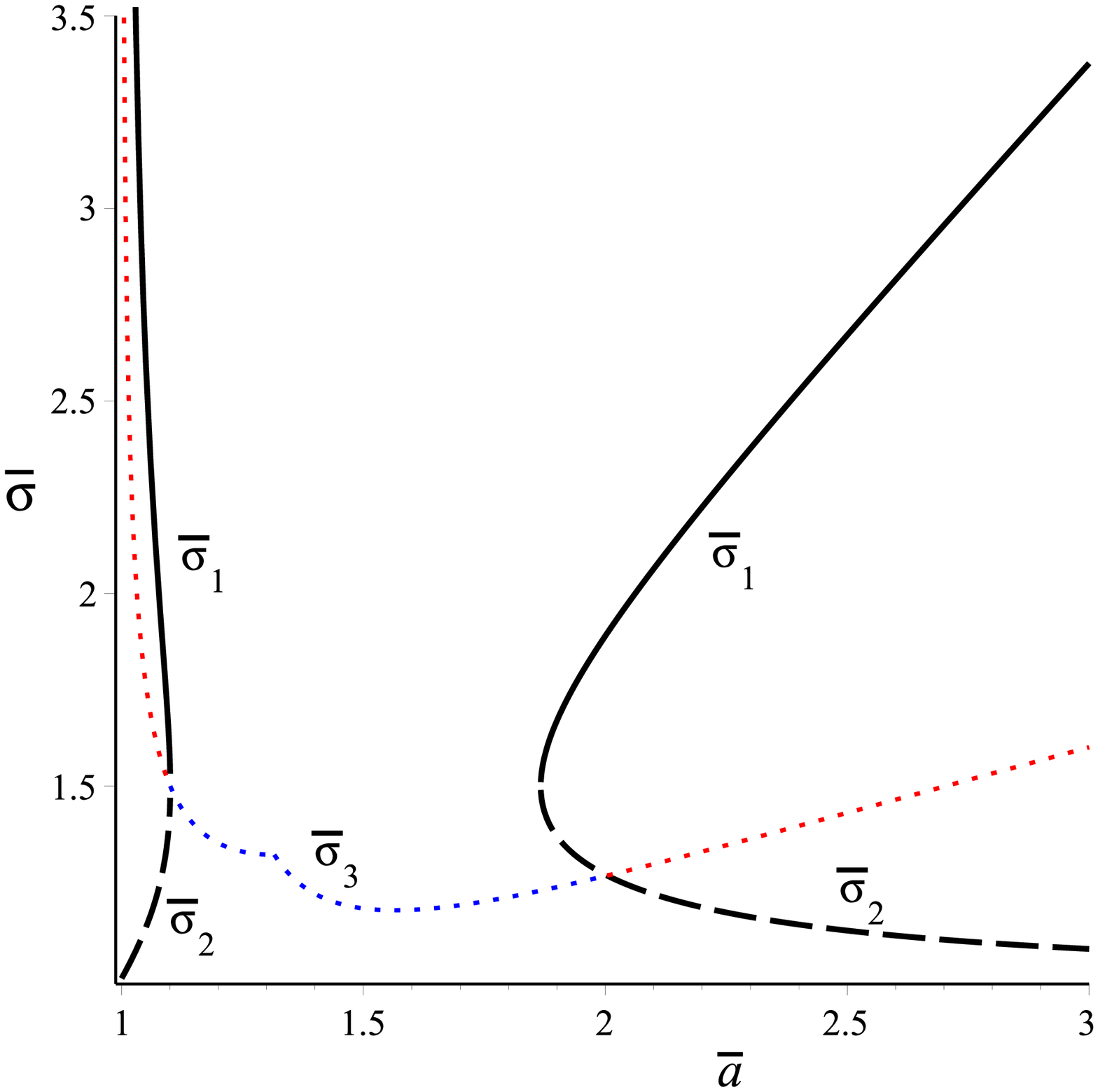}
\caption{Illustrative values $B_s=0.40$, $B_s=(B_s)_c\approx0.62$ and $B_s=0.75$, depicting the borders of the solutions of Eq.~(\ref{maeda_cond1a}). The regions allowed by these lines are the ones above $\bar{\sigma}_1$ and below $\bar{\sigma}_2$ in the left-bottom and middle-bottom panels; in the right-bottom panel, the intermediate region between the black lines is the acceptable one, without taking into account the flare-out. The boundary of the flare-out condition Eq.~(\ref{maeda_cond2c}) is also presented in the bottom panels as dotted lines, and the region which lies below such lines are the acceptable ones. Clearly, the flare-out here is compound of two parts, depending on the sign of the coefficient multiplying $\zeta^0$ in Eq.\ (\ref{maeda_cond2c}). Thus, the portion of the flare-out line in between the two solutions also constitutes a third solution border \(\bar{\sigma}_3\), which is presented in blue color in the middle-bottom and right-bottom panels and as a dotted line (also in color online) in the corresponding conformal diagram (the middle-top and right-top panels). Note that all the lines depicted in these conformal diagrams are (excluded) borders of the throats.}
\label{maeda_conf}
\end{figure}
\end{widetext}

The substitution of the equations of motion for $\bar a$ and $\bar \sigma$ into Eq.\ (\ref{maeda_cond2c}) yields
\begin{equation}
\label{maeda_cond2d}
\left[\frac{B_s^2}{\bar a^2}\left(\frac{3}{\bar a^4}-1\right)\right](\zeta^0)^2+\left[\frac{1}{2\bar\sigma^3}+\frac{B_s^2}{\bar a^2}\left(\frac{1}{\bar a^4}-1\right)\right](\zeta^1)^2>0,
\end{equation}
Since the coefficient multiplying $\zeta^0$ in the ``flare-out'' condition (\ref{maeda_cond2d}) depends only on $\bar a$ and it vanishes at $\bar a=\sqrt[4]{3}$, we can split the analysis of such equation into two parts: if $\bar a\leq\sqrt[4]{3}$, then we can set $\zeta^0=0$ in Eq.\ (\ref{maeda_cond2d}) in order to minimize the positive contribution of $\zeta^0$. Thus, the flare-out will be satisfied for any $\zeta^A$ if the coefficient multiplying $\zeta^1$ in Eq.\ (\ref{maeda_cond2d}) is positive. On the other hand,  if $\bar a>\sqrt[4]{3}$, then we use the limiting case $\zeta^0=\zeta^1$ (i.e., light-like vectors) in Eq.\ (\ref{maeda_cond2d}) in order to maximize the negative contribution of such component. In this case, the flare-out becomes

\begin{equation}
\label{maeda_cond2e}
\bar a^6-4B_s^2\left(\bar a^4-2\right)\bar\sigma^3>0,
\end{equation}
which corresponds to the minimality condition of the HV's and SH's definitions, as one would expect.

One can see from Fig.\ \ref{maeda_conf} that the wormhole solution comprehends the region interior to the trapping horizon of SH's definition when $B_s$ is smaller than its critical value, as well as the static limit is achieved for $B_s\longrightarrow0$. For values of $B_s$ equal or greater than its critical number, there are some portions of the throats, in particular the central regions of the conformal diagrams, for which the contour lines of the throats lie outside the local light cones. Therefore, any time-like observer at those regions has no choice on entering or leaving the throat. Finally, it is worth noticing that according to MHC definition the space-time regions where the wormhole throat is defined do not necessarily violate the NEC condition. Nevertheless, these regions in our solution do violate the NEC condition, as we shall see afterwards.

\subsection{Tomikawa, Izumi and Shiromizu's definition}\label{defTIS}

Based on the aforesaid definitions \cite{hoch98a,hay99,mae09}, Y. Tomikawa, K. Izumi and T. Shiromizu (TIS) provided a hybrid definition \cite{tomi15} for the wormhole throats, which is also constructed from null congruences on space-like hypersurfaces. However, it turns out to be independent of the choice of the foliation and it is more intuitive according to the authors. When singularities are present, the TIS definition assumes that the NEC and the cosmic censorship conjecture hold in order to avoid trivial situations. If at least one of these hypotheses is not valid, black holes or FLRW space-times could be seen as wormholes, for instance. Nevertheless, the cosmological model we are dealing with here is singularity free, due to the violation of the strong energy condition, then this restriction do not apply to our case.

The TIS definition considers the null geodesic congruences of a given manifold parameterized by the affine parameters $u_{\pm}$ and a co-dimension two compact surface $\Sigma$ on it, such that the following two quantities are defined from the expansion coefficients associated with each congruences:
\begin{equation}
k=\Theta_{+}-\Theta_-,
\end{equation}
and
\begin{equation}
\bar{k}=\Theta_{+}+\Theta_-.
\end{equation}
With the definition of the auxiliary vector $r^{\mu}=(\p/\p u_{+}-\p/\p u_{-})^{\mu}$, the characterization of the throat $\Sigma$ is
\begin{equation}
\label{tomi1}
k\big|_{\Sigma}=0
\end{equation}
and the flare-out condition is
\begin{equation}
\label{tomi2}
r^{\mu}\nabla_{\mu}k\big|_{\Sigma}>0.
\end{equation}
For the same congruence of null geodesics used before, these equations yield in our case
\begin{equation}
\label{theta_tomi}
\frac{\sigma_\chi}{\sigma}=0,\quad \mbox{and} \quad \frac{\sigma_{\chi\chi}}{\sigma}>0.
\end{equation}
\begin{figure}[!htb]
\centering
\includegraphics[width=5cm,height=5cm]{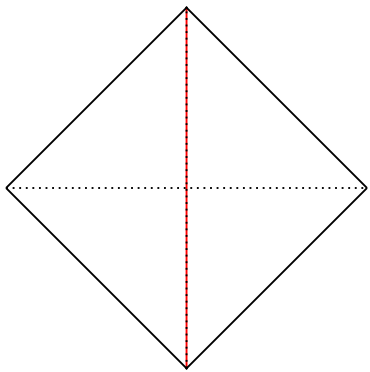}
\caption{Wormhole throat (vertical solid line) in the TIS definition. The same conformal diagram holds for any value of the scale parameter associate to the magnetic field.}
\label{tomi_conf}
\end{figure}
Remarkably, this result is independent of the time evolution of the universe. Therefore, this is the weakest condition for a wormhole in a bouncing cosmology. With the help of Eqs.~(\ref{cons_R}) and (\ref{first_sigma}), Eq.~(\ref{theta_tomi}) can be rewritten as
\begin{equation}
\label{theta_tomi2}
\sigma=2M,\quad \mbox{and} \quad \sigma>0,
\end{equation}
which are trivially satisfied. It means that the throat is univocally defined at the minimum radius for all time. Note that the scale parameter $B_s$ does not appear in the equations that define the throat, implying that this definition provides the same profile for all models. Furthermore, the conformal diagram in this case fits the ones of a static wormhole (Morris-Thorne \cite{mor88} and Ellis \cite{ellis73}), therefore the throat can be crossed as many times as desired.

\section{Energy conditions}\label{energy}

\begin{figure}[htb]
\flushleft
\vspace{1.8cm}
\includegraphics[width=11.5cm,height=16cm]{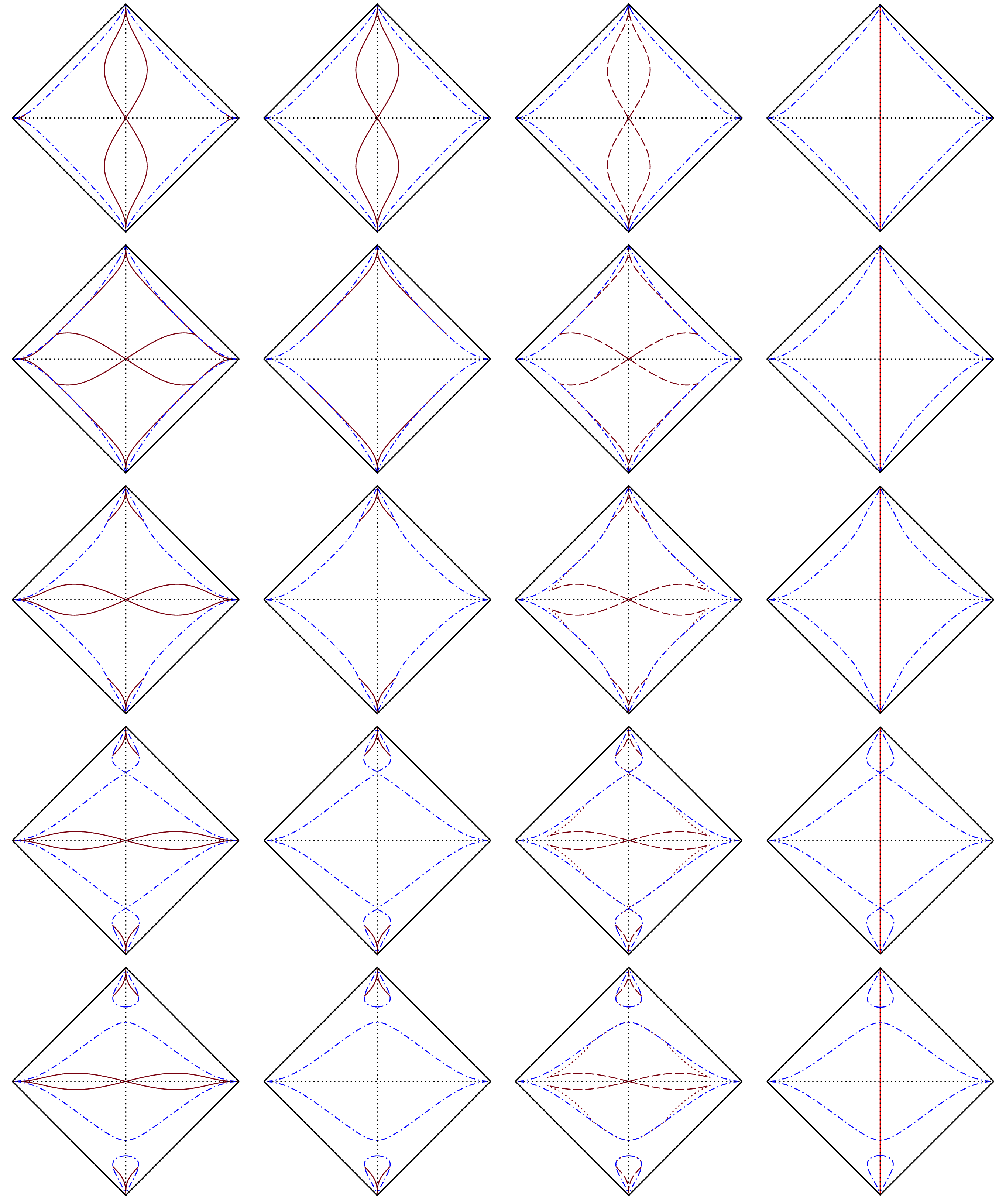}
\caption{The conformal diagrams displaying HV's definition are the panels shown on the left, with SH's diagrams in the second column, MHC's ones in the third, and TIS's ones on the right. The boundaries of the NEC, Eq.~(\ref{nec2}), were included as dot-dashed lines (blue online) with illustrative values of the scale parameter $B_s=0.40$, $B_s=(B_s)_c$, $B_s=0.75$, $B_s=\sqrt[4]{27/32}\approx0.96$ and $B_s=1.00$, respectively in the rows from top to bottom. Note that in all cases the wormhole is formed only where the NEC is violated, except for TIS's definition.}
\label{final-nec}
\end{figure}

Let us now focus on the energy conditions and their eventual violation when a wormhole forms. We begin with the null energy condition, which states that should not be negative the total projection of the energy-momentum tensor \(T^{\mu\nu}\) along any light-like vector \(k_\mu\), {\em i.e.},
\begin{equation}
\label{nec}
\rho_n:=T^{\mu\nu}k_\mu k_\nu\geqslant0.
\end{equation}
With the help of the null property \(k^\mu k_\mu=0\), then Eq.~(\ref{Tmunu}) inserted in the NEC, with a diagonal space-like anisotropic pressure tensor \(\pi^\mu{}_\nu\) yields \(\rho_n=(\rho+p-\pi^1{}_1/2)k^0k_0+3\pi^1{}_1k^1k_1/2\). The ratio \(-(k^1k_1)/(k^0k_0)\) obviously lies in the range \([0,\,1]\). Therefore, the NEC in the model being considered reads
\begin{equation}
\label{nec2}
\rho+p-\pi^1{}_1\geqslant0.
\end{equation}
It should be clear from Eq.~(\ref{nec2}) that a NEC violation does not necessarily require an exotic fluid configuration, since the anisotropic pressure \(\pi^1{}_1\geqslant0\) contributes with a negative sign. That is to say, NEC violation is expected to occur for non-negative \(\rho\) and \(p\) if the anisotropic pressure \(\pi^1{}_1\) dominates among these three quantities. Note that such situation is actually expected to take place in the asymptotic future of an expanding universe with either flat or open spatial sections and equation of state \(p=\omega\,\rho\) with \(\omega>-1/3\) (since \(p\sim\rho\sim a^{-3(1+\omega)}\) and \(\pi^1{}_1\sim a^{-2}\)). 
Furthermore, from the analysis of Eq.~(\ref{nec2}) written in terms of $\bar a$ and $\bar \sigma$, there is a bifurcation for the NEC violating regions occurring at $B_s=\sqrt[4]{27/32}$.

It is illustrative to insert the NEC boundaries in the conformal diagrams of the wormholes, as shown in Fig.\ \ref{final-nec}. For HV's definition, most of the results are rather expected, since the appearance of the throats formally demands NEC violation. The plots are provided as the left panels in Fig.~\ref{final-nec}. It is apparent that a sufficiently large value of the mean magnetic field, here encoded into the scale parameter \(B_s\), allows room for localized throats in the asymptotic past and future regions, which are dominated by the presence of the anisotropic pressure, while there are limited epochs in which the universe satisfies NEC everywhere. From the diagrams of the second column in Fig.~\ref{final-nec}, it is straightforward to see that SH's definition is indeed a particular case of HV's one. The temporal condition removes the portion of the trapping horizon which is space-like, leading to two-way traversible surfaces solely. The situation changes a bit when MHC's definition is being considered, since the region where the wormhole lies do not require NEC violation (as it was already stated by the authors, even though in lack of any definite example). The results are also shown as the third column in Fig.~\ref{final-nec} for easier comparison, with a range of values of the magnetic field for which the wormhole can be detected in regions which do not satisfy NEC. It is worthwhile to remind the reader that MHC defines the wormhole as the interior region delimited by the dashed lines. For TIS's definition, the diagrams look somehow trivial. They are also shown as the right panels in Fig.~\ref{final-nec} for an easy comparison of the four definitions.


\section{Summary and final remarks}\label{two-way}

In all the cases dealt with, the mouths (or throats depending on the definition) appear in pairs with causal connection with one another. It is apparent from the plots in Fig.~\ref{final-nec} that HV's definition yields, for the present solution, a wormhole throat portion which does not correspond to our intuition of what a wormhole should look like. This is because of the portions of the throat far away from the center, which exist for any value of the magnetic field and eventually dominates (and extends to everywhere) for sufficiently strong magnetic fields. Actually, such portions do correspond to traversable surfaces, but causality forbids traversing them backwards. That is to say, such portions are one-way traversable, instead of being a two-way traversable surface ({\em i.e.}, back and forth as one pleases) as it is expected to be the case for regular wormholes. This criticism was originally raised by Hayward in physical terms \cite{hay99} and partially put in mathematical form later on \cite{hay09}. Thus, HV's proposal would require further restrictions in order to capture the intuitive meaning of a wormhole.

For MHC's definition, the situation is a bit more involved. While being possibly uncovered by NEC violation regions, its one-way traversability occurs only for large values of the magnetic field. Thus, this proposal yields the expected result for weak magnetic fields, and becomes less well-behaved for strong fields. That is to say, the claim for generalization of the notion of a wormhole may be disguised by working in the limit of small fields only.

For TIS's proposal, the resulting throat lies exactly at the same place as in the corresponding static Morris-Thorne solution \cite{mor88}. Such proposal gives a wormhole throat which, while also being uncovered by NEC violation regions, is everywhere two-way traversable. It is worth noticing that NEC violation does not require an exotic matter content in our case, as it can be induced solely by the presence of anisotropic pressures as shown in Sec.~\ref{energy}. In principle, TIS's definition could appear as the most reliable one for a wormhole. Notwithstanding, it does not take into account the dynamics of the universe, which we would expect to affect the global structure of the throats.

SH's definition, roughly speaking, simply erases from HV's diagrams the non-temporal portions of the mouths, thus providing two-way traversability by construction. Among all four definitions, SH's one gets the closest to what a simple minded poor guy pictures for a wormhole. However, for large fields the SH's mouths are not the borders of a spacetime region in the two-dimension conformal diagrams shown above, since these mouths do not close in the diagram. Of course, this phenomenon is an effect of the whole dynamics, due probably to the energy interchange between the wormhole and the bounce along the time. This means that SH's definition may yield two mouths of a wormhole with no ``tunnel'' in between. Such structure resembles Lewis Carol's Cheshire's cat (a rather uncommon cat which smiles, the particularity of which being that the smile of the cat can exist with no cat lying around it).

In conclusion, a generalization of an already known exact solution of Einstein field equations \cite{bitt14} was here interpreted as being a non-stationary wormhole solution. The first integrals of the geodesic equations are calculated and the causal structure of this solution is enlightened by its Carter-Penrose conformal diagram which is rather similar to the Minkowski space-time. Then, we apply the quasi-local definitions of a dynamical wormhole found in the literature to this particular solution and search for its peculiarities in terms of conformal diagrams, bifurcations in the free parameters and violation (or not) of the NEC. Finally, a crude comparison between the four different proposals appears to favor SH's definition \cite{hay99,hay09}, while HV's \cite{visser96,hoch98b}, MHC's \cite{mae09}, and TIS's \cite{tomi15} proposals demand revision based upon the concept of what a dynamic wormhole is physically expected to be.


\end{document}